\documentclass[fleqn,usenatbib]{mnras}

\usepackage[T1]{fontenc}

\DeclareRobustCommand{\VAN}[3]{#2}
\let\VANthebibliography\thebibliography
\def\thebibliography{\DeclareRobustCommand{\VAN}[3]{##3}\VANthebibliography}

\usepackage{graphicx}	% Including Figurefiles
\usepackage{amsmath}	% Advanced maths commands
\usepackage{amssymb}	% Extra maths symbols
\usepackage{soul}
\usepackage{newtxtext,newtxmath}
\usepackage{subcaption}
\usepackage{graphicx}
\usepackage{threeparttable}
\usepackage[font=small,skip=0pt]{caption}
\usepackage{float}
\usepackage[export]{adjustbox}
\title[Simulated host galaxy analogs of high-z quasars observed with JWST]{Simulated host galaxy analogs of high-z quasars observed with JWST}

\author[S. Berger et al.]{Sabrina Berger,$^{1, 2}$\thanks{E-mail: sabrina.berger@student.unimelb.edu.au}
Madeline A. Marshall,$^{2,3}$
J. Stuart B. Wyithe,$^{1, 2,4}$, 
Tiziana di Matteo$^{5}$, 
Yueying Ni$^{6}$,\newauthor and Stephen M. Wilkins$^{7}$
\\
$^{1}$School of Physics, University of Melbourne, Parkville, VIC 3010, Australia\\
$^{2}$ARC Centre of Excellence for All Sky Astrophysics in 3 Dimensions (ASTRO 3D). Australia\\
$^{3}$National Research Council of Canada, Herzberg Astronomy \& Astrophysics Research Centre, 5071 West Saanich Road, Victoria, BC V9E 2E7, Canada\\
$^{4}$Research School of Astronomy and Astrophysics, Australian National University, Canberra, ACT 2611, Australia\\
$^{5}$McWilliams Center for Cosmology, Department of Physics, Carnegie Mellon University, Pittsburgh, PA 15213, USA\\
$^{6}$Center for Astrophysics–Harvard $\&$ Smithsonian, 60 Garden Street, Cambridge, MA 02138, USA\\
$^{7}$Astronomy Centre, University of Sussex, Falmer, Brighton BN19QH, UK\\
}

\date{Accepted XXX. Received YYY; in original form ZZZ}

\pubyear{2023}

\begin{document}
\label{firstpage}
\pagerange{\pageref{firstpage}--\pageref{lastpage}}
\maketitle

\begin{abstract} 
The hosts of two low-luminosity high-z quasars, J2255+0251 and J2236+0032, were recently detected using JWST's NIRCam instrument. These represent the first high-z quasar host galaxy stellar detections and open a new window into studying high-z quasars. We examine the implications of the measured properties of J2255+0251 and J2236+0032 within the context of the hydrodynamic simulation BlueTides at z = 6.5. We find that these observed quasars fall on the BlueTides stellar to black hole mass relation and have similar luminosities to the brightest simulated quasars. We predict their star formation rates, estimating approximately $10^{2-3}$ $M_{\odot}/ \rm yr$ for both quasar hosts. J2255+0251 and J2236+0032's host galaxy radii also fall within estimates of the radii of the simulated host galaxies of similar luminosity quasars. We generate mock JWST NIRCam images of analogs to the observed quasars within BlueTides and perform a point source removal to illustrate both a qualitative and quantitative comparison of the measured and simulated radii and magnitudes. 
The quasar subtraction works well for similar luminosity quasars, and the recovered host images are consistent with what was observed for J2255+0251 and J2236+0032, further supporting the success of those observations.
We also use our mock imaging pipeline to make predictions for the detection of J2255+0251 and J2236+0032's hosts in upcoming JWST observations. We anticipate that the simulation analogs of future high-z quasar host discoveries will allow us to make accurate predictions of their properties beyond the capabilities of JWST.
\end{abstract} 


%%%%%%%%%%%%%%%%%%%%%%%%%%%%%%%%%%%%%%%%%%%%%%%%%%

%%%%%%%%%%%%%%%%% BODY OF PAPER %%%%%%%%%%%%%%%%%%

\section{Introduction}
% overview of quasars and high-z quasars
% Thousands of quasars have been discovered, mostly at $z \lesssim 2$ in surveys such as the Sloan Digital Sky Survey (SDSS) \citep{2002AJ....123.2945R}. 
Quasars are a subset of active galactic nuclei (AGN) or galaxies with highly accreting central supermassive black holes (SMBHs). A number of surveys have been dedicated to discovering high-z quasars. The Subaru High-z Exploration of Low-Luminosity Quasars \citep[SHELLQs;][]{Matsuoka_2016, Matsuoka_2018, Matsuoka_2019, Matsuoka_2019b, Matsuoka_2022}, the Canada-France High-z Quasar Survey (CFHQSIR) \citep{Pipien_2018}, SCUBA2 high-z Bright Quasar Survey (SHERRY) \citep{Li_2020}, the Dark Energy Spectroscopic Instrument (DESI) survey \citep{2019AJ....157..236Y, Yang_2020, Wang_2021, yang2023desi}, and Sloan Digital Sky Survey (SDSS) \citep{Jiang_2016} have discovered a significant number of quasars with $z \gtrsim 5$. Recently \citealt{matsuoka2023quasar} published the first quasar luminosity function at z = 7. The most recent SDSS survey of $z = 5.7-6.4$ quasars \citep{Jiang_2016} and the SHELLQs survey of $z = 5.7-6.9$ quasars \citep{Matsuoka_2018} span a luminosity range of $-29.0 ~ \rm mag < {M}_{1450} < -24.5~ \rm mag$ or $6.6 \times 10^{12} L_{\odot} < {L}_{\rm bol} < 3.8 \times 10^{14} L_{\odot}$ and $-26 ~ \rm mag < {M}_{1450} < -22~ \rm mag$ or $7.3 \times 10^{11} L_{\odot} < {L}_{\rm bol} < 2.5 \times 10^{13} L_{\odot}$, respectively. The existence of these quasars (with $M_{\rm BH}$ greater than $10^9 M_{\odot}$ in some cases) at such early times presents a challenge to models of galaxy formation (\citealt{Haiman_Loeb_2001,Latif_Schleicher_Schmidt_Niemeyer_2013}; \citealp[for a review see][]{Inayoshi_Visbal_Haiman_2020}).

In the nearby Universe, there are tight correlations between galactic properties such as stellar mass, black hole mass (i.e., the Magorrian relation, see \citealt{Magorrian_1998, Croton_2006, Schutte_2019}), bulge mass (\citealt{2013ARA&A..51..511K, Reines_2015}), velocity dispersion (\citealt{2000ApJ...539L...9F, Munari_2013, Sharma_2021}), size, and luminosity (\citealt{kormendy_1977, Danieli_2019}). The relationship between SMBHs and their host galaxy properties has been studied extensively at low redshifts due to their abundance and proximity. In the local Universe, direct measurements of black hole mass using reverberation mapping have allowed measurements of galactic scaling relations such as the Magorrian relation (\citealt{bentz_2015, Bentz_2018}).

% observations of quasars
Prior to the launch of the James Webb Space Telescope (JWST), observations of high-z quasar host galaxies had only been attainable at sub-mm wavelengths with instruments such as the Atacama Large Millimeter/sub-millimeter Array (ALMA) \citep{Schleicher_2010, Venemans_2012, Wang_2013, 2016ApJ...816...37V, Decarli_2018, Yue_2021}. These ALMA observations were successful in detecting [CII] emission, a strong tracer of the dusty interstellar medium. The rest-frame far-infrared emission observed at sub-mm wavelengths only traces the cold dust of the quasars' host. The rest-frame UV emission observed in the near infrared is a tracer for the stellar light of a quasars' host but requires careful subtraction of the quasar's light. At high-z, black hole mass measurements have been made by combining single epoch emission lines such as MgII and CIV and continuum emission to quantify the velocity and radius of the broad line region of the quasar \citep{2002MNRAS.337..109M}. Attempts were made to observe $z \sim 6$ quasar hosts with the Hubble Space Telescope (HST), resulting in upper limits on the host galaxy's brightness and stellar mass \citep{Mechtley_2012, McGreer_2014, 10.3847/1538-4357/abaa4c}. Despite accurate point spread function (PSF) modeling and careful quasar subtraction, the quasar hosts were not detected. Using the BlueTides simulation, \citet{10.48550/arxiv.2206.08941} predicted that approximately $50\%$ of high-z quasar hosts would be detectable with an equivalent JWST filter and exposure time. 

Measurements of detected high-z quasar hosts will allow the precise extension of galactic scaling relations to high-z through direct stellar mass measurements. This will extend our knowledge of the co-evolution of black holes and their host galaxies providing vital information into how the first galaxies formed. It will also shed light on the accelerated growth of SMBHs in early galaxies and their formation histories. 

Recently, the hosts of two low-luminosity high-z quasars: SHELLQs J2255+0251 and SHELLQs J2236+0032 at $z\simeq6.4$ with rest-UV absolute magnitudes of $-23.9~\rm mag$ and $-23.8~ \rm mag$, respectively, were detected with higher resolution JWST observations \citep{10.48550/arxiv.2211.14329}. The sources J2255+0251 \citep{Matsuoka_2018} and J2236+0032 \citep{Matsuoka_2016} were discovered in the SHELLQs survey amongst more than 150 other high-z quasars \citep{Matsuoka_2022}. The host galaxy of the slightly more luminous quasar, J2255+0251, was detected in just one JWST Near Infrared Camera (NIRCam) filter: F356W (3.135-3.981 $\mu \rm m$). J2236+0032's host was detected in both the F356W and F150W (1.331-1.668 $\mu \rm m$) filters. 

The deconvolving of the quasar from its host galaxy's stellar light requires careful fitting of the point source and galactic light profile. This ensures an accurate estimation of parameters such as the Sérsic magnitude and radius. \citet{10.48550/arxiv.2211.14329} used the \texttt{galight} software \citep{2020ApJ...888...37D} to decouple the stellar light from the quasar by modeling the quasar as a two dimensional PSF and Sérsic profile of the host galaxy. The Sérsic profile is defined as 
\begin{equation}
\label{eq:sersic}
    I(R) = I_e \exp \left \{-b_n \left[\left(\frac{R}{R_{1/2}}\right)^{1/n} - 1\right]\right\},
\end{equation}
where $R_{1/2}$ is the half light radius we define as the Sérsic radius, n is the Sérsic index corresponding to increasing steepness in the log brightness and log radius space, $b_n \sim 2n - \frac{1}{3}$, and $I_e$ is the intensity at the half light radius. This is done separately in both the F356W and F150W filters. This was successful except for J2255+0251 which was only detected in F356W. 

Soon after, three other high-z quasar hosts were detected in the JWST Emission-line galaxies and Intergalactic Gas in the Epoch of Reionization (EIGER) project. EIGER is a JWST Guaranteed Time Observation (GTO) program combining NIRCam and its accompanying deep wide-field slitless spectroscopy (WFSS) instrument. Part of EIGER's first observations have been published detailing observations of 6 quasars between redshifts 5.9 and 7.1 \citep{yue2023eiger}. The quasars were imaged in the NIRCam filters F115W, F200W, and F356W with wide-field slitless spectroscopy from F356W. Three out of six of these quasars' host galaxies were detected \citep{yue2023eiger}. 
The properties of the EIGER quasars with detected hosts, J0148+0600, J159-02, and J1120+0641, are found similarly except using \texttt{psfmc} \citep{2014PhDT.........1M}. From the PSF subtraction and fit, the UV magnitude, galactic radius, Sérsic index, ellipticity, and stellar mass were also measured (See tables \ref{tab:quasar_prob} and \ref{tab:eiger} for an overview of the properties of these quasars).

\citet{stone2023detection} also detected the host of J2239+0207 at $z \simeq 6.25$ by scaling the PSF to match the quasar flux and finding a discrepancy in the difference between the two. We do not include this detection's measurements in our paper to stay consistent with detections that are performed with Bayesian point source removal software, i.e., \texttt{psfmc} \citep{2014PhDT.........1M} and \texttt{galight} software \citep{2020ApJ...888...37D}. With non-Bayesian methods, \citet{stone2023undermassive} set upper limits on the galaxy flux from five high-z quasars that were \textit{not} detected with JWST. These quasar host detections and non-detections have provided a first glimpse into measuring the black hole to stellar mass relationship in high-z quasars. Although \citet{10.48550/arxiv.2211.14329}'s black hole mass measurements seem to be in line with the low-z relation, many other observations have measured overmassive black holes \citep{yue2023eiger,stone2023undermassive}. For example, the EIGER quasars are reported to have black hole to stellar mass ratios that are up to 2 dex higher than the local relation \citep{yue2023eiger}.

As more high-z quasar hosts are detected with JWST, questions about the scatter of key high-z black hole scaling relations have arisen. The errors from the sensitive point source subtraction, calibration uncertainties, and other instrumental effects are difficult to disentangle from the observed properties. This motivates the use of large scale high-z simulations and mock imaging techniques to quantify this bias and explore discrepancies. Other similar simulations do not have large enough volumes to accurately capture high-z quasar formation. For example, the large scale hydrodynamic simulation Illustris \citep{Vogelsberger_2014} is nearly 45 times smaller in volume compared to BlueTides. The subsequent simulation to BlueTides, ASTRID, is approximately 4 times smaller in size \citep{Bird_2022}. Due to its large box size, BlueTides presents itself as the ideal simulation to study high-z quasars with luminosities high enough (at least at the faint end of the observed quasar luminosity function) to be observable with current telescopes in a statistically robust way.

In this paper, we examine the measured properties of J2255+0251, J2236+0032, and the EIGER quasars with detected hosts within the distribution of $z \sim 6.5$ galaxies with black holes simulated in the BlueTides simulation. In section \ref{sec:methods}, we provide a brief description of the BlueTides simulation. In section \ref{sec:results}, we describe the specifics of converting the observed JWST parameters to comparable quantities in BlueTides. We then quantify our comparison between these quasars and the predictions of the BlueTides simulation. In section \ref{sec:mock_images}, we describe our methods for generating mock images and spectra of the BlueTides analogs of J2255+0251 and J2236+0032 and also performing a point source removal. We compare the results from the mock spectra, images, and point source removal in section \ref{sec:comparing_mocks}. We make predictions for quasar host detectability in upcoming observations of the quasar analogs with the followup filters and exposure times in section \ref{sec:predict}. Lastly in section \ref{sec:conclusions}, we summarize our conclusions and provide future outlooks on upcoming JWST quasar measurements. We assume a $\Lambda \rm CDM$ cosmology with parameters from WMAP9 (\citealt{Hinshaw_2013}) to as follows, $H_0= 100~\rm (km/sec)/Mpc \times \textit{h} = 67.4 (km/sec)/Mpc$, $\Omega_m=0.2814$, $\Omega_b=0.0464$, $\Omega_{\Lambda}=0.7186$, $\sigma_8=0.820$, and $n_s=0.9710$. We express all magnitudes in AB magnitude.

\begin{table*}\centering  	
\caption{Relevant parameters \tnote{*} for J2255+0251 and J2236+0032 as measured or determined from \citet{10.48550/arxiv.2211.14329}, or in their original discovery papers: \citet{Matsuoka_2016} and \citet{Matsuoka_2018}. All black hole errors are MCMC random errors and do not account for the usual 0.4 dex \citep{2006ApJ...641..689V} systematic errors from the scatter in the empirical relation used in \citet{10.48550/arxiv.2211.14329}. \label{tab:quasar_prob}} 
\begin{threeparttable}
	\centering
	\resizebox{\textwidth}{!}{
 \begin{tabular}{ccccccccccc} 
		\hline
	Quasar & z & $M_{\rm 1450}^{\rm QSO}$ [mag] & $m_{\rm F150W}^{\rm QSO}$ [mag] & $m_{\rm F356W}^{\rm QSO}$ [mag] & $L_{\rm bol} [L_{\odot}]^{*}$ & $L_{\rm bol}/L_{\rm Edd}^{*}$ & $\dot{M} [M_{\odot}/yr]^{*}$ & $\log_{10}$ $M_{*}~[M_{\odot}]$ & $M_{\rm BH}~[M_{\odot}]$ & $R_{\rm 1/2}~[\rm kpc]$ \\ \hline
    J2236+0032 & 6.40 & $-23.66 \pm 0.10$ & $22.78 \pm 0.02$ & $21.75 \pm 0.02$ & $3.12 \times 10^{12}$ & 0.07 & 2.13 & $11.12^{0.40}_{-0.27}$ & $13.6 \pm 1.5 \times 10^{8}$ & $0.5 \pm 0.2$ (F150W); $0.7 \pm 0.1$ (F356W) \\ \hline
    J2255+0251 & 6.34 & $-23.87 \pm 0.04$  & $22.20 \pm 0.02$ & $22.03 \pm 0.02$ & $3.76 \times 10^{12}$ & 0.59 & 2.56  & $10.53^{+0.51}_{-0.37}$ & $1.97 \pm 0.17 \times 10^{8}$ & $1.5 \pm 1.1$ (F356W) \\ \hline
\end{tabular}}
\begin{tablenotes}\footnotesize
\item[*] Parameters with an $^*$ were calculated from observed parameters using methods described in section \ref{subsec:conversion_lbol}.
\end{tablenotes}
\end{threeparttable}
\end{table*}

\begin{table*}

	\centering
\caption{Relevant parameters for the EIGER quasars with host galaxies detected as measured or determined from \citet{yue2023eiger}, or in their original discovery papers: \citet{Bañados_2016}, \citet{Yang_2021}, \citet{Farina_2022}, \citet{10.1093/mnras/stad1468}, and \citet{Mazzucchelli_2023}. All black hole errors are MCMC random errors and do not account for the usual 0.4 dex \citep{2006ApJ...641..689V} systematic errors from the scatter in the empirical relation used in \citet{yue2023eiger}. \label{tab:eiger}} 
\begin{threeparttable}
	\resizebox{0.98\textwidth}{!}{
 \begin{tabular}{cccccccccccc} 
		\hline
	Quasar & z & $M_{\rm 1450}^{\rm QSO}$ [mag] & $m_{\rm F115W}^{\rm QSO}$ [mag] & $m_{\rm F200W}^{\rm QSO}$ [mag] & $m_{\rm F356W}^{\rm QSO}$ [mag] & $L_{\rm bol} [L_{\odot}]^{*}$ & $L_{\rm bol}/L_{\rm Edd}^{*}$ & $\dot{M} [M_{\odot}/yr]^{*}$ & $\log_{10}$ $M_{*}~[M_{\odot}]$ & $\log_{10}M_{\rm BH}~[M_{\odot}]$ & $R_{\rm 1/2}~[\rm kpc]$ (F356W) \\ \hline
    J0148+0600 & 5.977 & -27.08 & $19.522 \pm 0.003$ & $18.912 \pm 0.001$ & $19.109 \pm 0.003$ & $6.65 \times 10^{13}$ & 0.26 & 45.2 & $10.74^{+0.31}_{-0.30}$ & $9.892^{+0.053}_{-0.055}$ & $2.23 \pm 0.11$ \\ \hline
    J159-02    & 6.381 & -26.47                             & $20.146 \pm 0.003$                  & $19.680 \pm 0.002$                  & $19.543 \pm 0.003$                  &  $3.84 \times 10^{13}$                         &      0.94                     &    26.1                      & $10.14^{0.34}_{-0.36}$                      & $9.096^{0.007}_{-0.005}$      & $2.64 \pm 0.16$       
\\ \hline J1120+0641 & 7.085 & -26.44                             & $20.366 \pm 0.003$                  & $19.886 \pm 0.002$                  & $19.632 \pm 0.003$                  &  $3.74 \times 10^{13}$                         &      0.96                     &        25.4                  & $9.81^{0.23}_{-0.31}$                       & $9.076^{0.029}_{-0.030}$      & $1.66 \pm 0.10$        
 \\ \hline
\end{tabular}}
\begin{tablenotes}\footnotesize
\item[*] Parameters with an $^*$ were calculated from observed parameters using methods described in section \ref{subsec:conversion_lbol}.
\end{tablenotes}
\end{threeparttable}
\end{table*}

\section{The BlueTides simulation}
\label{sec:methods}
\subsection{Simulation Overview}
\label{sec:bluetides}
BlueTides is a hydrodynamic simulation run between z = 99 and z = 6.5 \citep{Feng_2015, Tenneti_2018, 2020MNRAS.495.2135N} using the MP-Gadget Smoothed Particle Hydrodynamics (SPH) code. The simulation contains $2 \times 7040^3$ particles in a box with sidelength = $400h^{-1}\rm Mpc$ and includes star formation, feedback processes, and black hole accretion in its computed physics. It contains 200 million star-forming galaxies \citep{Huang_Matteo_Bhowmick_Feng_Ma_2018}. Owing to the finite volume of BlueTides, the most massive black hole is $7.7 \times 10^8 M_{\odot}$ at z = 7 and $1.4 \times 10^9 M_{\odot}$ at z = 6.5. Higher black hole masses are unlikely to be created within the $400h^{-1}\rm Mpc$ box. Previous studies have shown good agreement between BlueTides and observational constraints on the high-z Universe, including quasar number counts \citep{Tenneti_Matteo_Croft_Garcia_Feng_2017, 2020MNRAS.495.2135N} and galaxy stellar mass relationships \citep{Wilkins_Feng_Matteo_Croft_Lovell_Thomas_2017}. All these previous studies were at z = 7. We present the first analysis of BlueTides at z = 6.5. 

The BlueTides AGN formation and feedback model was developed in \citet{2005Natur.433..604D}. The MassiveBlack I and II simulations \citep{Di_Matteo_2012, 10.1093/mnras/stv627} share the same AGN formation prescription as BlueTides. To align with predictions from the direct collapse scenarios for black holes \citep{Yue_2014}, BlueTides uses a black hole seed mass of $10^{5.8} M_{\odot}$. They occur within haloes that are larger than approximately $10^{10.8} M_{\odot}$. To avoid studying black holes that are too close to the black hole seed mass, we limit our study to 108,000 black holes which have the highest black hole accretion rate at z = 6.5. These black holes are hosted within virial halos with a minimum mass of $10^{11.0} M_{\odot}$, maximum mass of $10^{13.1} M_{\odot}$, and average mass of $10^{11.4} M_{\odot}$. The maximum black hole mass for those not included in the first 108,000 galaxies is approximately $10^{6.3} M_{\odot}$. Thus, we limit our sample to include black holes with masses greater than $10^{6.3} M_{\odot}$ when finding JWST analogs within BlueTides. This BlueTides subset includes black holes with bolometric luminosities (calculated using equation \ref{eq:BH_acc}) less than $10^{14.0} L_{\odot}$ hosted within galaxies with stellar masses less than $10^{11.3} M_{\odot}$. 

\begin{figure}
\includegraphics[width=\columnwidth]{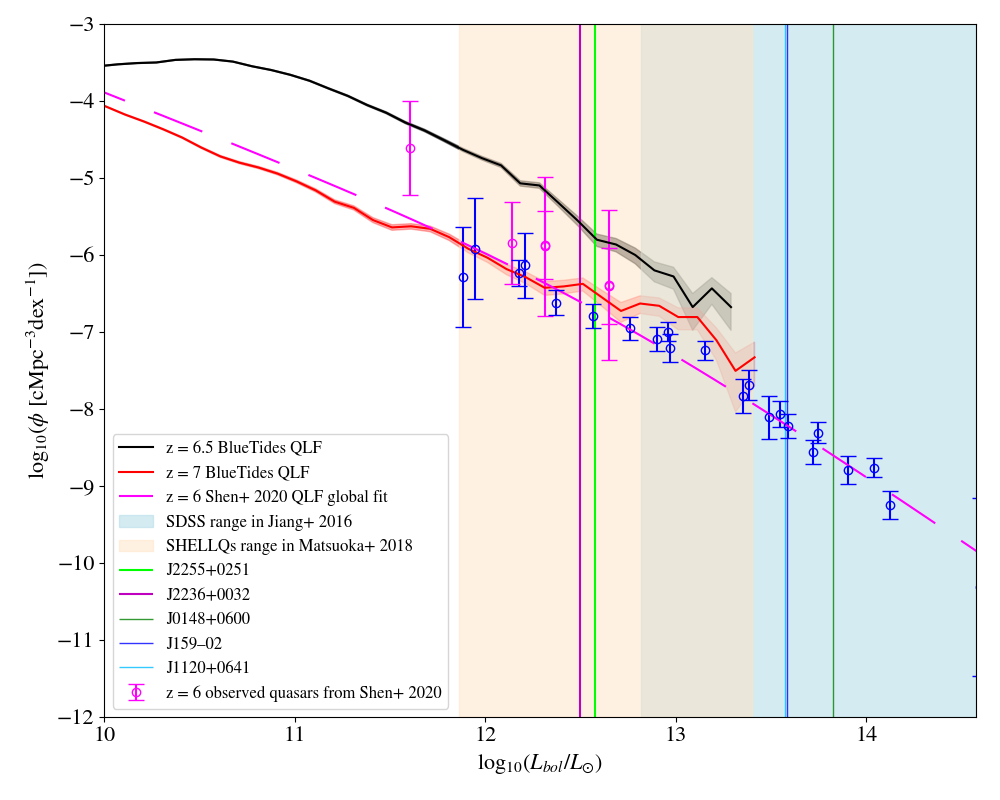}
    \caption{The BlueTides bolometric quasar luminosity function (QLF) is plotted for all black holes at $z = 6.5$ in black with Poisson errors in shaded grey. Similarly in red with errors in pink, the $z = 7$ BlueTides QLF is plotted for comparison. The \citet{10.48550/arxiv.2211.14329} quasars have bolometric luminosities similar to the brightest quasars in BlueTides. We also overplot the range of bolometric luminosity depth for two quasar surveys: the SHELLQs and SDSS quasars with $z=5.7-6.9$ from \citet{Matsuoka_2018} and \citet{Jiang_2016} in tan and light blue, respectively. The z = 6 global quasar luminosity function from \citet{10.1093/mnras/staa1381} is overplotted in blue for reference with the observations from the fit. The fit used is referred to as global fit A in \citet{10.1093/mnras/staa1381} where the lower luminosity portion of the fit is a flexible polynomial. Our smallest bins contain one quasar.}
    \label{fig:quasar_pdf}
\end{figure}

\subsection{Quasar Luminosity Function (QLF)}
In Figure \ref{fig:quasar_pdf}, we show the z = 7 BlueTides QLF plotted alongside the z = 6.5 QLF. We compare this to observed quasars, including the parameterized QLF from \citet{10.1093/mnras/staa1381}, which is overplotted for reference (global fit A in \citealt{10.1093/mnras/staa1381}). It is described by the following single power law: 
\begin{equation}
       \phi_{\rm bol}(L) = \frac{\phi_{*}}{2(L/L_{*})^{\gamma}},
\end{equation}
where $\gamma = 1.509 \pm 0.058$, $\log \phi_{*} = -5.452$, and $\log L_{*} = 11.978 \pm 0.055$. Since the $z \sim 6$ quasar luminosity function does not include many low luminosity quasars (with UV luminosities greater than about -21 mag), the typical double power law used to describe the quasar luminosity function is replaced with a single power law to avoid degeneracies in the parameter fit. 

Very few quasars have been found above z = 7, and at z = 6.5, BlueTides provides a sample of quasars that is more representative of those that have been discovered so far. We see that BlueTides produces a realistic quasar population, with quasars matching the lowest known luminosity quasars and spanning the lower end of the range of SHELLQs quasars. Due to the finite volume of BlueTides, the BlueTides QLF does not extend far into the brightest SDSS quasars or include many quasars matching the luminosity of the EIGER quasars. We also find that the lower redshift QLF peaks an order of magnitude higher in quasar bolometric luminosity as expected for older black holes.

% In Figure \ref{fig:glf}, we show that this calibration still suits the simulation at z = 6.5. 
% \subsection{UV Galaxy Luminosity Function (GLF)}
% Since we present the first application of the BlueTides simulation at z = 6.5, we compare the UV galaxy luminosity functions (GLF) at z = 6.5 and z = 7. \citet{10.1093/mnras/stx841} found that for BlueTides galaxies with z > 8, the galaxy stellar mass function and far-UV luminosity function match observations. We compare the simulated GLFs to the observed GLFs between $z \sim 6.5$ and $z \sim 7$ from \citet{2015ApJ...803...34B} in Figure \ref{fig:glf}. We find that our dust model calibrated to the observed z = 7 GLF is suitable for the z = 6.5 BlueTides galaxies as described in section \ref{subsec:mock_images_methods}.

% \begin{figure}
% \includegraphics[width=1.15\columnwidth]{plots/GLF_comparison.png}
%     \caption{The BlueTides UV galaxy luminosity function (GLF) is plotted for the brightest galaxies at z = 7 and z = 6.5. The intrinsic GLFs are plotted in dashed lines, and the attenuated GLFs are plotted in solid lines. We see that the dust correction z = 7 can be transferred to z = 6.5. The crossing of the GLFs is due to the number of galaxies in both the dust attenuated and intrinsic GLFs being constant. Our smallest bin contains 5 galaxies.}
%     \label{fig:glf}
% \end{figure}

\section{Comparing quasars to BlueTides}
\label{sec:results} 
\subsection{Black hole bolometric and Eddington luminosity}
\label{subsec:conversion_lbol}
To identify mock quasars within BlueTides with the properties of observed high-z quasars, we use the mass accretion rate of the black hole to calculate its corresponding bolometric luminosity. We can then calculate their corresponding UV magnitudes, which is accomplished using a fit from measured quasar spectral energy distributions (SEDs). 

The UV magnitudes of J2255+0251 and J2236+0032 were previously measured in the SHELLQs survey. The absolute UV quasar magnitudes are $-23.87 \pm 0.04$ and $-23.66 \pm 0.10$ for J2255+0251 and J2236+0032, respectively. For the EIGER quasars, the UV magnitudes for the quasars are -27.08, -26.47, and -26.44 for J0148+0600, J159-02, and J1120+0641, respectively.  As described in \citet{qin_2017}, we can convert a quasar bolometric luminosity to a magnitude in UV at 1450 \r{A} using \citet{Hopkins_2007} extrapolated from the B-band 4344 \r{A} magnitude using \citet{2007AJ....133..734B},

\begin{equation}
\label{eq:bol}
    M_{\rm UV, 1450} = M_{\rm bol} + 2.5 \log_{10} k_{\rm UV, 1450} - 0.434,
\end{equation}
where
\begin{equation}
\label{eq:corr}
k_{\rm B} \equiv \frac{L_{\rm bol}}{L_{\rm B}}=6.25 \left(\frac{L_{\rm bol}}{10^{10}L_{\odot}}\right)^{-0.37} + 9.00 \left(\frac{L_{\rm bol}}{10^{10}L_{\odot}}\right)^{-0.012}.
\end{equation}
There is no analytic form for the inverse of equation \ref{eq:corr}. We solve numerically for the bolometric luminosity, which was calculated for all quasars explored in this work and is tabulated in \ref{tab:quasar_prob}.
% \begin{equation}
% \label{eq:inv}
% 10^{-M_{\rm UV} + 4.74}(x^{-2.5}k_{\rm UV}^{2.5}) = 1,
% \end{equation}

From the black hole accretion rate, we approximate $L_{\rm bol}$ of the black hole with
\begin{equation}
    \label{eq:BH_acc}
    L_{\rm bol} = \eta \dot{M}_{\rm BH}c^2,
\end{equation}
where $\dot{M}_{\rm BH}$ is the quasar mass accretion rate and $\eta$ is the radiative efficiency.
We take $\eta = 0.1$ assuming an equipartition magnetic field \citep{Gruzinov_1998}. We compare the bolometric luminosity to the Eddington luminosity of a black hole, which is
\begin{equation}
L_{\rm edd} = \frac{4 \pi G M_{\rm BH} c}{\kappa} = 1.26 \times 10^{38}\left(\frac{M_{\rm BH}}{M_{\odot}}\right) \rm erg s^{-1},  
\end{equation}
where $\kappa$ is the opacity of the black hole. If we assume the highly accreting material around a black hole to be mostly ionized hydrogen so that Thomson scattering describes the opacity reasonably well, we can set $\kappa = \sigma_{\rm T} / m_{\rm p}$, where $\sigma_{\rm T}$ is the Thomson cross section and $m_{\rm p}$ is the mass of a proton. We note that Bondi (or spherical) accretion is assumed in BlueTides. 

We estimate the bolometric to Eddington luminosity ratio \citep{Hopkins_2007} from the measured black hole masses in \citet{10.48550/arxiv.2211.14329} and \citet{yue2023eiger}. The Eddington luminosities are $6.4 \times 10^{12} L_{\odot}$ and $4.9 \times 10^{13} L_{\odot}$ for J2255+0251 and J2236+0032, respectively. Using their estimated bolometric luminosities, the bolometric to Eddington luminosity ratio is approximately 59$\%$ and 7$\%$ for J2255+0251 and J2236+0032, respectively. For the EIGER quasars, the Eddington luminosities are $2.54 \times 10^{14} L_{\odot}$, $4.06 \times 10^{13} L_{\odot}$, and $3.88 \times 10^{13}$ for J0148+0600, J159-02, and J1120+0641, respectively. With the estimated bolometric luminosities, the Eddington ratio is approximately $26\%$, $94\%$, and $96\%$ for J0148+0600, J159-02, and J1120+0641, respectively. This puts the luminosities of J159-02 and J1120+0641 near the Eddington limit.

\subsection{Black hole mass}
Figure \ref{fig:quasar_pdf} shows that with luminosities near $3 \times 10^{12} L_{\odot}$, J2255+0251 and J2236+0032 fall within the most luminous quasars in BlueTides as described in section \ref{sec:bluetides}. This allows us to compare other properties of the observed quasars with analogs in BlueTides. The EIGER quasars are in the BlueTides luminosity range where there are only a handful of quasars. With so few quasars, the binned luminosity function does not extend to such a high luminosity as seen in Figure \ref{fig:quasar_pdf}.

Figure \ref{fig:bh} shows the \textit{simulated} black hole masses for quasars in BlueTides with the \textit{measured} quasar masses overplotted. Both \citet{10.48550/arxiv.2211.14329} quasars' measured black hole mass and luminosity fall within the expected BlueTides distribution. The quasars have tight errorbars and fall within the expected black hole mass range. J2236+0032 has a black hole mass above that predicted by BlueTides for quasars with similar bolometric luminosities. The EIGER quasars J159-02 and J1120+0641 also fall within the expected black hole mass range, but the most massive quasar, J0148+0600, has a mass of $10^{9.892}~M_{\odot}$ which is above the mass limit of BlueTides. 
\subsection{Stellar mass and SFR}
\begin{figure*}
    \centering
    \begin{subfigure}[t]{\columnwidth}
    \includegraphics[width=1.2\columnwidth]{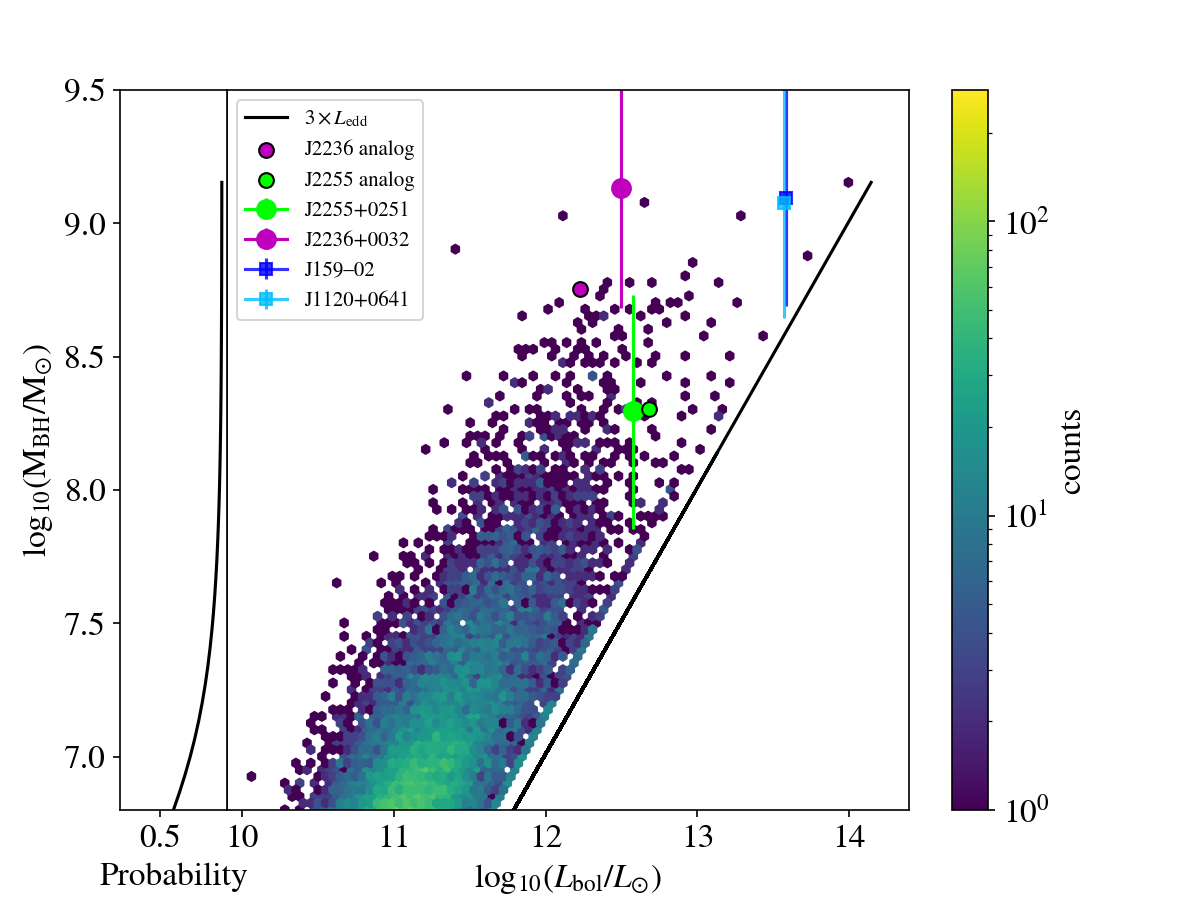}
    \caption{}
    \label{fig:bh}
    \end{subfigure}
   \begin{subfigure}[t]{1\columnwidth}
    \includegraphics[width=1.2\columnwidth]{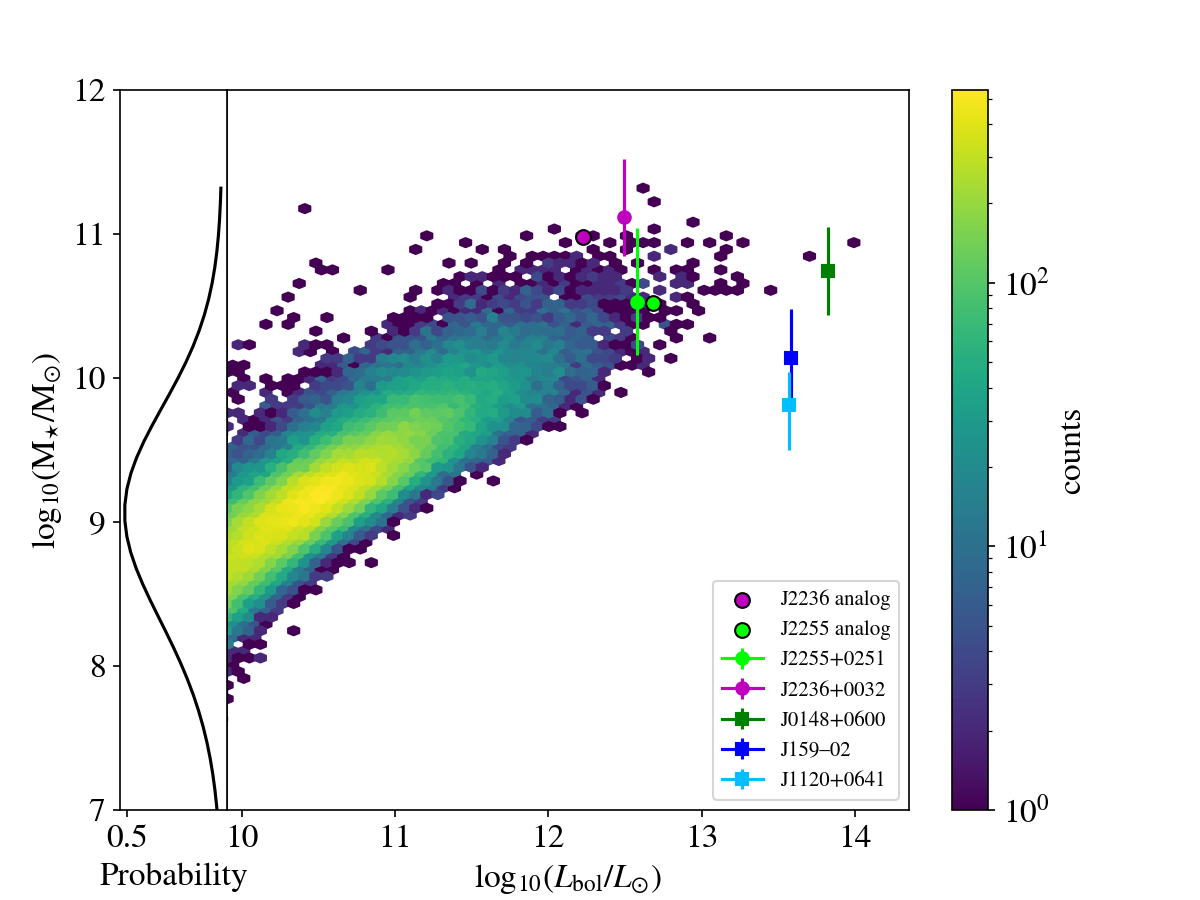}
    \caption{}
    \label{fig:stellar_mass_sub_1}
    \end{subfigure}
    \caption{Each plot displays a combination of stellar mass, black hole mass, and bolometric luminosity for the BlueTides black holes and the J2255+0251, J2236+0032, and EIGER \citep{yue2023eiger} quasar properties overplotted. The J2255+0251 and J2236+0032 BlueTides analogs (see section \ref{subsec:jwst_analog_selection}) are plotted in green or purple for J2255+0251 and J2236+0032, respectively with a black circle. Left: Bolometric luminosity and black hole mass for simulated and observed quasars. The observed quasar mass is the measured value from NIRSpec observations. The black line indicates the luminosity limit for BlueTides quasars which is approximately 3 times the Eddington luminosity. Right: The same as above but for stellar mass versus bolometric luminosity. \label{fig:original}}
\end{figure*}

\begin{figure*}
    \centering
    \begin{subfigure}[t]{\columnwidth}
    \centering
    \includegraphics[width=1.2\columnwidth]{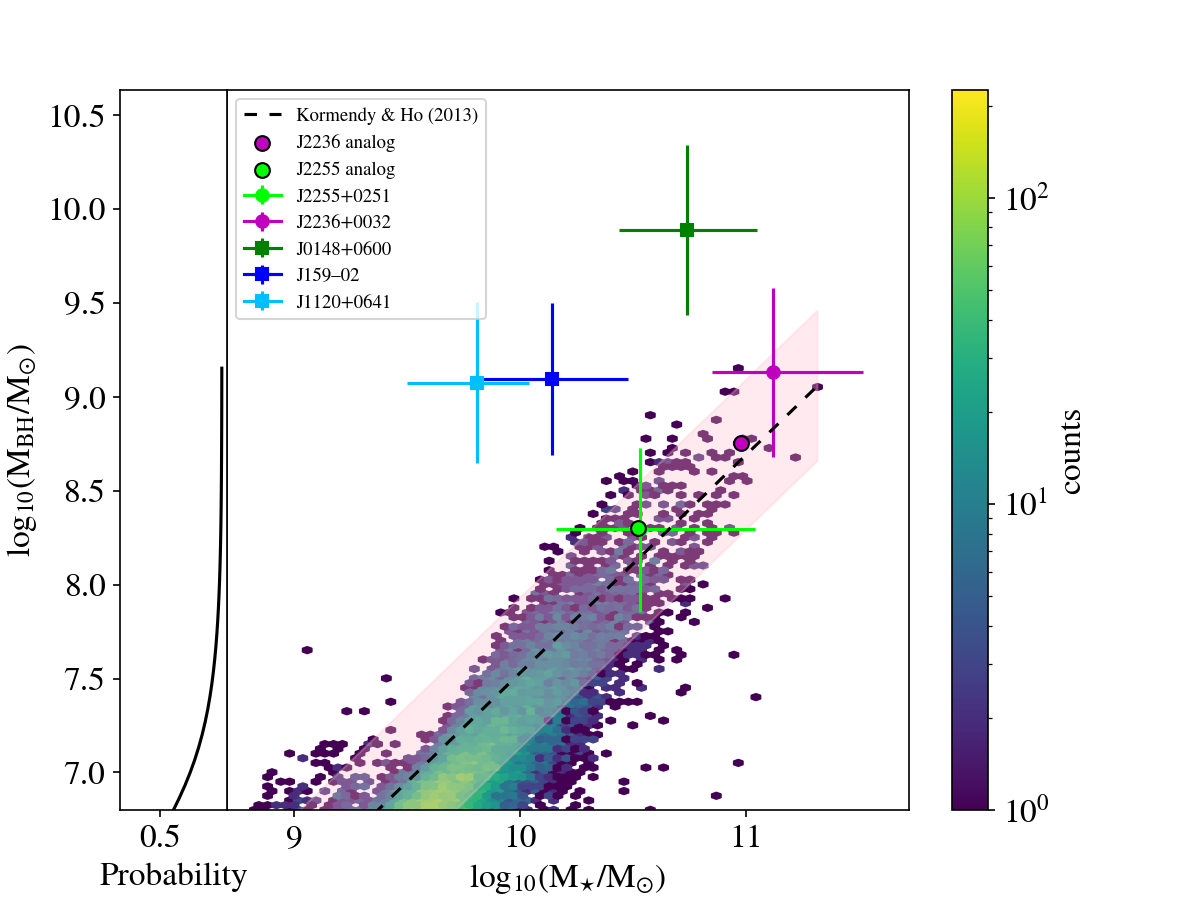}
    \caption{}
    \label{fig:stellar_mass_sub_3}
    \end{subfigure}
    \begin{subfigure}[t]{\columnwidth}
    \centering
    \includegraphics[width=1.2\columnwidth]{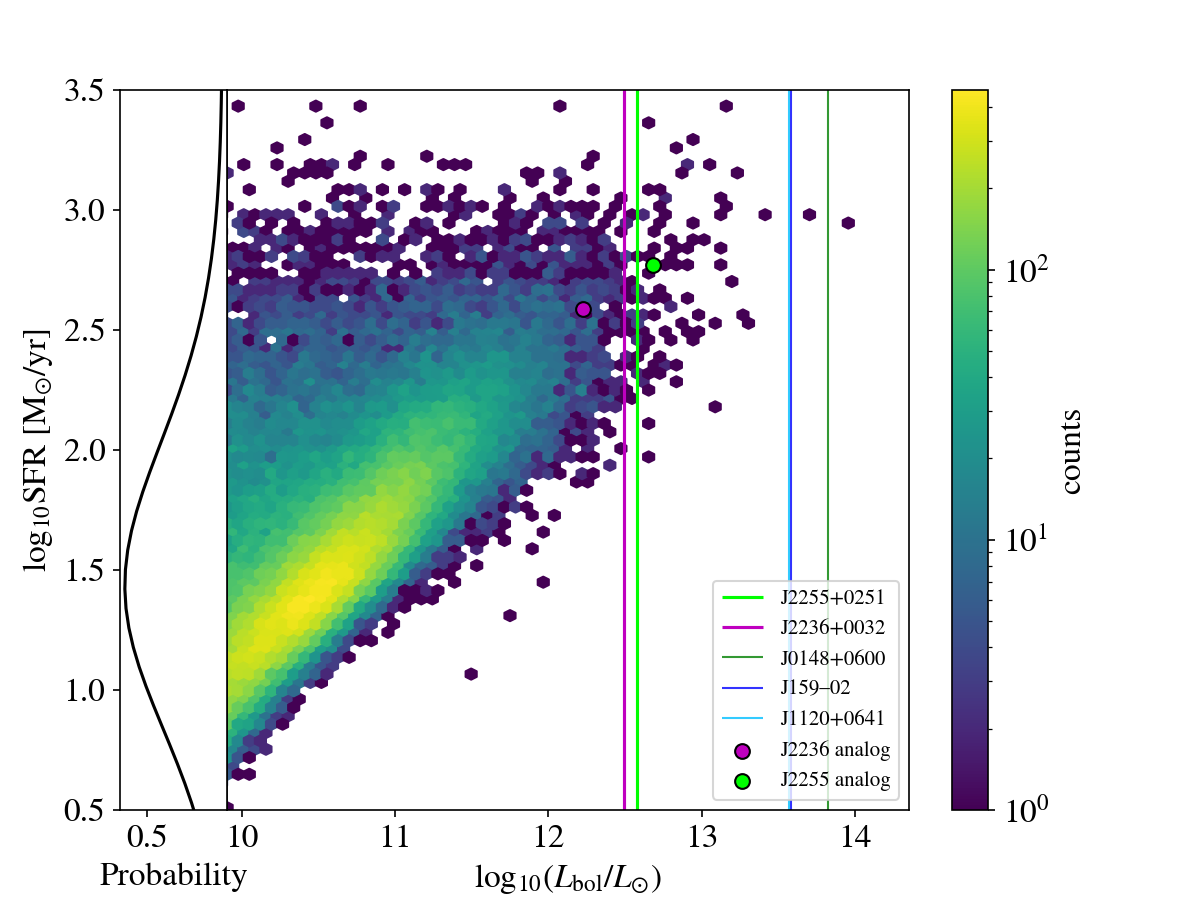}
    \caption{}
    \label{fig:sfr}
    \end{subfigure}
    \caption{Left: The same as Figure \ref{fig:original} but for black hole versus stellar mass, i.e., the Magorrian relation. We overplot the local black hole mass to bulge mass relation as described in \citet{2013ARA&A..51..511K} in the dashed black line along with the systematic error of 0.4 in light pink. Right: Same as Figure \ref{fig:original} for SFR and bolometric luminosity of BlueTides galaxies.}
\end{figure*}

Figures \ref{fig:stellar_mass_sub_1} through \ref{fig:stellar_mass_sub_3} show the joint distribution of black hole masses and luminosity with stellar mass. The stellar mass was measured as $10^{10.53} M_{\odot}$ and $10^{11.12} M_{\odot}$ for J2255+0251 and J2236+0032, respectively \citep{10.48550/arxiv.2211.14329}. The black hole masses and stellar masses of J2255+0251 and J2236+0032 follow the black hole mass to stellar mass ratio. However, the EIGER quasars lie 1-2 dex above the \citet{2013ARA&A..51..511K} relation and do no lie within the BlueTides distribution. The EIGER quasars have bolometric luminosities that are brighter than the BlueTides quasars which may bias the comparison. The uncertainty and difficulty of removing the quasar from its host (as described in section \ref{subsec:mock_images_methods}) could lead to biased stellar mass measurements. Understanding this observational bias will be the goal of future work.
% For BlueTides mock quasars within 50$\%$ of J2255+0251 and J2236+0032's bolometric luminosities, the measured hosts fall within the expected range of host stellar masses. This masked version includes 68 quasars and can be seen more explicitly in the panels \ref{fig:stellar_mass_sub_2} and \ref{fig:stellar_mass_sub_4}. 

In Figure \ref{fig:sfr}, we show where the quasars would lie on the star formation rate versus bolometric luminosity distribution. This allows us to estimate their SFR using the BlueTides distribution of quasars. We anticipate a star formation rate between approximate $10^{2-3}$ $M_{\odot}/ \rm yr$ for both quasars. For the more luminous EIGER quasars, we can place an approximate lower limit of $10^{2.75}$ $M_{\odot}/ \rm yr$. Followup measurements in the sub-mm with instruments such as ALMA of the H$\alpha$ emission line with JWST will directly measure the SFR. This shows the potential of using high-z simulations to both constrain and support future quasar observations. 

\subsection{Galaxy radii}
\label{sec:radii}
\begin{figure}
    \centering
     \includegraphics[width=1.1\columnwidth]{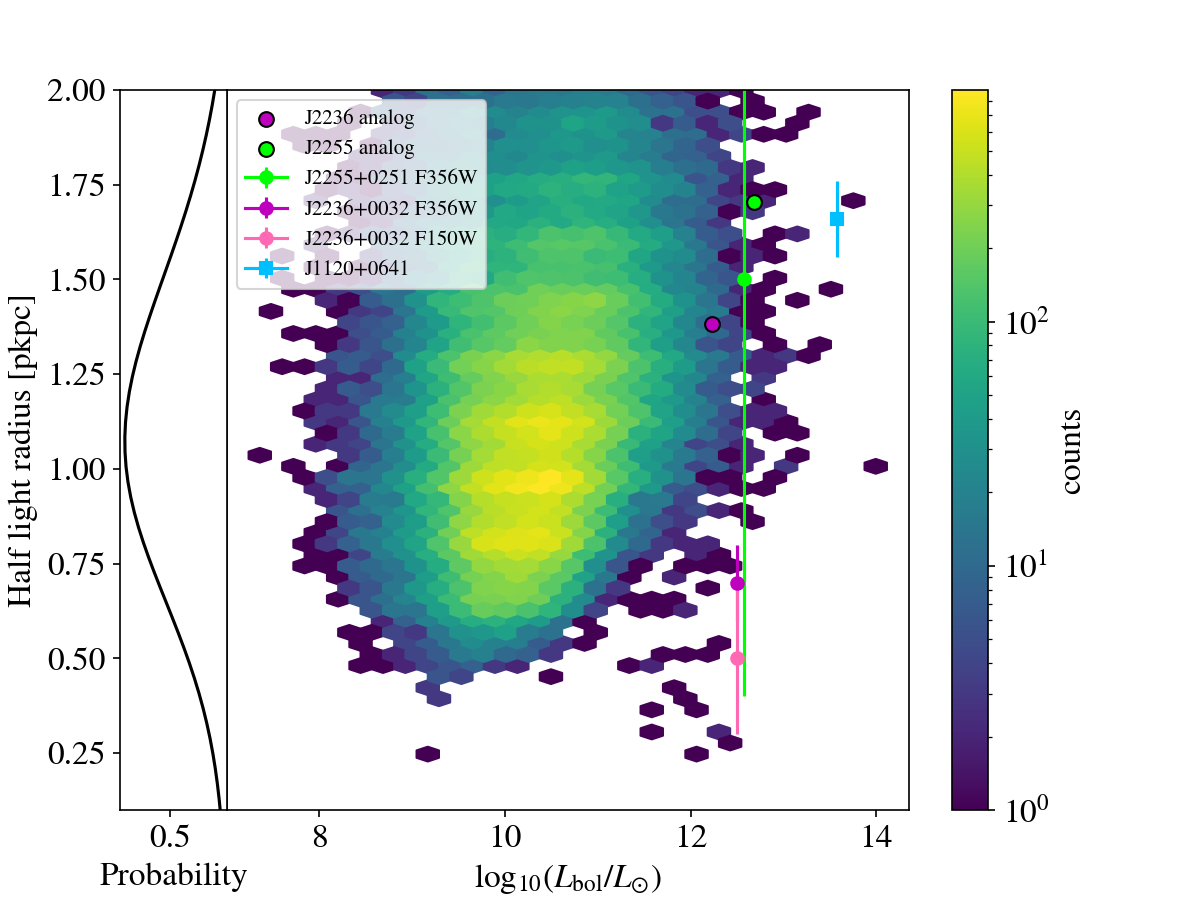}
    \caption{Same as Figure \ref{fig:original} but for the half light radius versus bolometric luminosity for BlueTides galaxies. The measured properties of recently detected high-z hosts overplotted.}
    \label{fig:rad_prop_bol}
\end{figure}

The half light radius ($R_{1/2}$) of J2255+0251, J2236+0032, and the EIGER quasars are estimated in \citet{10.48550/arxiv.2211.14329} and \citet{yue2023eiger} (where they are called effective radii). $R_{\rm 1/2}$ is the Sérsic half light radius measured by \texttt{galight} or \texttt{psfmc} which accounts for a deconvolution of the PSF. J2255+0251 is only detected in one filter (F356W) and has only one estimate for $R_{1/2}$. These values are tabulated in Table \ref{tab:quasar_prob} and Table \ref{tab:eiger} and range between approximately 0.5 to 3 kpc. 

In Figure \ref{fig:rad_prop_bol}, we compare the half light radii estimates of BlueTides galaxies from \citet{marshall_2022_dust} against their bolometric luminosity. These radii were calculated by generating rest-frame images in the standard FUV top hat filter (1500 \r{A}). They do not account for a PSF or any noise. The orientation and size of the field of view were found to make a difference less than 0.04 dex and are thus neglected in our comparisons. To calculate these half light radii, \citet{marshall_2022_dust} performed aperture photometry with 20 circular apertures logarithmically spaced between 0.05 to 2 kpc and interpolated between them to find the half light radius. We find that the half light radii measured in BlueTides and those fit in \citet{10.48550/arxiv.2211.14329} are consistent as seen in Figure \ref{fig:rad_prop_bol}. For the EIGER quasars, we find that J1120+0641 is consistent with BlueTides. J159-02 and J0148+0600 have radii measurements of $2.64\pm0.1$kpc and $2.23 \pm 0.11$kpc, respectively, above our sample in radius.

In recent results with the JWST ERO SMACS 0723 Field for galaxies between redshifts 7 and 12 as seen in \citet{10.1093/mnras/stac3347}, the maximum half light radius measured is near 0.8 kpc. This result is in line with similar HST measurements found by \cite{Holwerda_2015} which measures sizes in the CANDELS survey. \cite{Holwerda_2015} measured an average radius size of $0.5 \pm 0.1$ kpc. Similarly in \citet{Yang_2022} which explore massive galaxies from the JWST GLASS survey with redshifts above 7, the half light radii found in the NIRCam filter F200W does not exceed $1.11 \pm 0.11 \rm kpc$ (Galaxy ID 4397). These radii values fall near the average of those found for BlueTides galaxies that host high bolometric luminosity quasars and recently detected quasar hosts from \citet{10.48550/arxiv.2211.14329}.

% \begin{figure}
%     \centering
%     \includegraphics[width=\linewidth]{plots/logged_bol_lum_half_radius_dust.png}
%     \caption{The half light radius versus bolometric luminosity for BlueTides galaxies with the measured properties of recently detected high-z hosts overplotted.}
%     \label{fig:rad_prop_bol}
% \end{figure}

In summary, the quasars hosts detected in \citet{10.48550/arxiv.2211.14329} are described well by the results of the BlueTides simulation and are in line with other high-z galaxy radii measurements. 

\section{Creating Mock Images and SPectra}
\label{sec:mock_images}
\subsection{BlueTides analog selection}
\label{subsec:jwst_analog_selection}
Finding the analogs of detected quasar hosts in BlueTides allows us to make predictions beyond what is possible with existing observations, and to assess the success of the simulation. We select the BlueTides analogs within the 4D property plane of black hole bolometric luminosity, black hole mass, stellar mass, and host galaxy radius. We find one analog per quasar within our joint distribution constraints. Each of these analogs is seen within the distributions shown in figures \ref{fig:bh} through \ref{fig:rad_prop_bol} in the same color as the respective observed JWST quasar with a black circle around the BlueTides galaxy. 

We adjust the range to explore for each property to ensure that we find at least one analog per observed quasar. Our analogs are within 0.4 dex of the measured bolometric luminosities for both quasars. They are within 0.4 and 0.01 dex of the measured black hole masses and 0.2 and 0.01 dex of the measured stellar masses for J2236+0032 and J2255+0251, respectively. The J2255+0251 BlueTides analog is within the error bars of the measured galaxy radius, but the J2236+0032 analog is about 0.7 kpc larger in galaxy radius than the observed host. In Figure \ref{fig:bh}, the J2255+0251 analog is seen with a black hole mass and bolometric luminosity nearly matching the observed quasar host. For J2255+0251, we choose the BlueTides galaxy with the closest properties as our analog, but for J2236+0032, we select the brightest quasar from approximately 5 analogs to remain closer to the magnitude of the detected host in \citet{10.48550/arxiv.2211.14329}. This allows for comparison of the point source subtraction between two analogs with varying magnitudes for J2236+0032. With a BlueTides galaxy analog for both J2236+0032 and J2255+0251, we are able to make comparisons and predictions between BlueTides and these high-z quasars.

% bluetides analogs
% stellar_mass, bh_mass
% J2236 analog
% [10.979563] [8.753221]
% J2255 analog
% [10.520594] [8.30382]

\subsection{Dust Calibration of BlueTides Star Particles}
\label{sec:dust}
A key component in comparing BlueTides to observations is ensuring accurate dust modeling during the mock imaging process described in section \ref{subsec:mock_images_methods}. Each star has a simple exponential dust model for the stellar natal cloud if stars are younger than 10 Myr. The optical depth of the natal cloud is calculated as follows

\begin{equation}
    \tau_{\rm birth~cloud} = 2 \left( \frac{Z}{Z_{\odot}} \right) \left( \frac{\lambda}{5500 \text{\AA}} \right)^{\gamma},
\end{equation}
where $Z$ is the metallicity of the birth cloud, $\gamma = -1$, and $\lambda$ is the wavelength at the particular optical depth. The interstellar medium (ISM) dust contribution is calculated for each star particle regardless of age as follows
\begin{equation}
    \tau_{\rm ISM} = \kappa \left(\frac{\lambda}{5500 \text{\AA}} \right)^{\gamma} \int_{d=0}^{d} \rho_{\rm metal}(x,y,z)~dz,
\end{equation}
where $\kappa = 10^{4.6}$, $\gamma = -1$, and $\rho_{\rm metal}$ is metal density in the line of sight as a function of $z$ which is integrated out to $d$, the distance to the star. $\gamma$ and $\kappa$ are calibrated from the observed UV galaxy luminosity function at z = 7 as shown in \citet{marshall_2022_dust}. 

\subsection{Mock images and spectroscopy}
\label{subsec:mock_images_methods}
% lead in to looking at mock images qualitatively
% qualitatively we see similar scales for these quasars
To compare model quasars directly with existing observations, we created mock JWST NIRCam images and processed these to recover host properties in analogy with observations. We generate mock images using the Python package, \texttt{SynthObs} \citep{synthobs} including both a mock background and simulated dust. The effects of adding both background and dust to a raw simulated galaxy image increase the galactic magnitude by $\sim 0.4$.

Within \texttt{SynthObs}, we assign a flux to each star particle based on each particle's metallicity, mass, and age using Binary Population and Spectral Synthesis (BPASS) \citep{Eldridge_2017, 2018MNRAS.479...75S} processed through Cloudy \citep{2017RMxAA..53..385F}. We select the BPASS model which uses the standard Salpeter stellar initial mass function (IMF) \citep{1955ApJ...121..161S} up to 300$M_{\odot}$. We assume 90$\%$ of Lyman-continuum photons escape. The dust model for each stellar particle is described in section \ref{sec:dust}. We then use the star particle flux and redshift to generate an SED for each star. We also add a point source to emulate a quasar from \citet{10.48550/arxiv.2211.14329} with the same observed apparent magnitude for each filter as shown in Table \ref{tab:quasar_prob}. Although we could model the SED of the BlueTides black hole in the analog galaxy, we choose to match the quasar magnitudes in the mock images to ensure a more similar point source removal process. The point source removal is highly dependent on the ratio of galaxy to quasar flux where a relatively brighter host typically allows for an easier host detection.

The filter transmission curves are convolved with each SED to get the appropriate filter specific flux for each particle. These SEDs get added together to form the overall spectrum of the galaxy. Finally, we convolve the spatial grid of filter corrected fluxes with mock JWST NIRCam PSFs from WebbPSF \citep{2015ApJ...798...68G}. For the higher resolution F150W filter, we do not incorporate any supersampling and generate a 25 x 25 kpc field with 0.031 arcsec/pixel with 144 pixels. The F356W filter's images are also generated with a 25 x 25 kpc field with 0.063 arcsec/pixel supersampled to 0.031 arcsec/pixel with 140 pixels. For each image at z = 6.5, we note that 1 pkpc = 0.18 arcsec. 

We also add background noise to the image to allow a more realistic fitting process. The background is derived from the aperture flux limit for each JWST filter at a 10$\sigma$ depth and the size of the aperture \citep{Rieke_2023}. The aperture flux limit for a particular filter is the minimum observable flux for a 10$\sigma$ observation of a point source for a specific exposure time. More explicitly, the background at each pixel for a given filter is
\begin{equation}
\label{eq:background}
    \rm background (\rm pixel) = \sqrt{(F_{limit}/10\sigma)^2 / A_{aperture}} \times N(0, 1),
\end{equation}
where $F_{\rm limit}$ is the aperture flux limit, $A_{\rm aperture}$ is the area of the aperture, and N is a random value drawn from a normal distribution at variance 1. We use an exposure time of 3.1ks to maintain consistency with \citet{10.48550/arxiv.2211.14329} for our mock point source removal, unless specified otherwise.

\section{Comparing BlueTides Mock Observations to the SHELLQs Quasars}
\label{sec:comparing_mocks}
\subsection{Mock Spectra of BlueTides Analogs}
We make redshifted mock spectra for the JWST analogs of J2255+0251 and J2236+0032 and 15 other close analogs with similar bolometric luminosities (some of their mock images can be seen in Figure \ref{fig:just_galaxies}). We infer the SED in the measured two-band photometry as described in section \ref{subsec:mock_images_methods}. The magnitudes and flux ratios for both the observed and simulated quasars are summarized in Table \ref{tab:quasar_mags}. Notably, the BlueTides host analogs have magnitudes broadly similar to the magnitudes of the observed quasar hosts. The upper limit for the host of J2255+0251 in F150W falls significantly below the magnitude of the BlueTides analog. For the filters where hosts were detected, this provides evidence that the primary analogs chosen are representative of the observed quasar hosts. For both observed quasar hosts, the spectra fit must be lower in F150W and higher in F356W to match the observed magnitudes. This may indicate that the observed host brightness was underestimated, due to the
difficulty of the quasar subtraction. 

\begin{figure*}
\centering
\includegraphics[width=\linewidth]{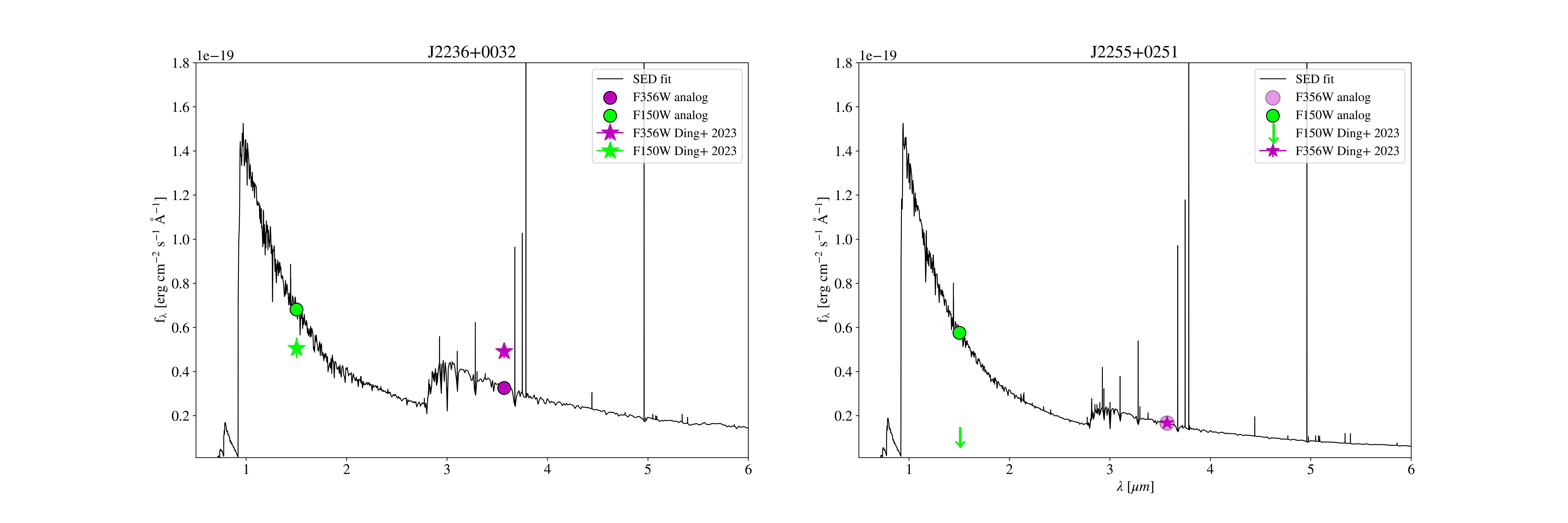}
    \centering
    \caption{Mock spectra from \texttt{SynthObs} of the two primary analog galaxies and 15 other galaxies hosting black holes with similar bolometric luminosities as those in \citet{10.48550/arxiv.2211.14329}. We also show the magnitudes of the primary analog and observed host in both F356W and F150W for both quasars for comparison. All mock spectra have similar shape but differ in amplitude due to stellar mass, stellar age, metallicity, and metal surface density.}
    \label{fig:finalspectra}
\end{figure*}

In Figure \ref{fig:finalspectra}, we plot the galaxy spectra for both the primary analogs and additional analogs to compare to Extended Data Figure 5 in \citet{10.48550/arxiv.2211.14329}. The mock spectra also look similar to high-z galaxy spectra observed with JWST in \citet{maiolino2023jwstjades} and \citet{2024NatAs.tmp...53B}. This indicates reasonably accurate spectral modeling with \texttt{SynthObs} of the analogs. 

Each analog has a similar spectral shape. The increasing amplitude of the spectra is roughly correlated with stellar mass but also depends on differences in average age, metallicity, and metal surface density for stars in the galaxies. All analog mock spectra also show the diversity of emission lines potentially present in the high-z quasar hosts which are much more difficult to detect in high-z quasar systems compared to the quasar spectra.

Although we do not perform a quantitative comparison, the spectra of the galaxy analogs look significantly different than the median template from SED inference in \citet{10.48550/arxiv.2211.14329}. Blueward of the Balmer break, the analogs show a much steeper spectra, instead being more consistent with their template corresponding to the 1$\sigma$ upper limits. This results in the significantly higher flux in both primary BlueTides analogs in the F150W filter than the observed hosts. We predict this difference in spectral shape between all analogs and the observed hosts to be primarily due to a difference in stellar age in modeled spectra. The median inferred spectra in \citet{10.48550/arxiv.2211.14329} uses a stellar population with an age of 370 Myr and 500 Myr for the galaxies of J2236+0032 and J2255+0255, respectively. For the primary BlueTides analogs, the average age is approximately 100 Myr resulting in a much steeper UV slope. However, the BlueTides galaxy spectra appear broadly consistent with the inferred spectra from the lower age range modeled in  \citet{10.48550/arxiv.2211.14329}.
% This different spectral shape likely implies a different age, metallicity, and/or dust attenuation for our galaxy than the median estimated by \citet{10.48550/arxiv.2211.14329}.

\begin{table*}
\resizebox{0.7\textwidth}{!}{
\begin{tabular}{cccccc}
\hline
Quasar               & Filter               & $m_{\rm Galaxy, JWST}$ [mag] & $m_{\rm Galaxy, sim}$  [mag] & JWST Observed Flux Ratio & Analog Flux Ratio \\ \hline
 J2255 &  F356W & $24.58 \pm 0.30$ & 24.2 & $9.8 \% \pm 2.6 \%$ & 12.5 $\%$ \\ \hline
 J2255 &  F150W & $> 26.3$ & 24.8 & $<3.8 \%$ & 14.7 $\%$ \\ \hline
 J2236 &  F356W & $23.12 \pm 0.20$ &  23.5 & $25.5 \% \pm 4.4 \%$ & 18.9 $\%$  \\ \hline
 J2236 &  F150W & $25.12 \pm 0.29$ &  24.6 &  $10.2 \% \pm 2.8 \%$ & 15.0 $\%$  \\ \hline
\end{tabular}}
\caption{The observed and analog galaxy magnitudes and ratios between galaxy and total (quasar + host) fluxes measured in \texttt{SynthObs}. The observed magnitudes and flux ratios are taken from Extended Data Table 1 in \citet{10.48550/arxiv.2211.14329} using errors on the found using the dispersion on the results from five different PSF models. \label{tab:quasar_mags}}
\end{table*}

\subsection{Mock Rest Frame Images of BlueTides Analogs}
\begin{figure*}
    \includegraphics[width=\linewidth]{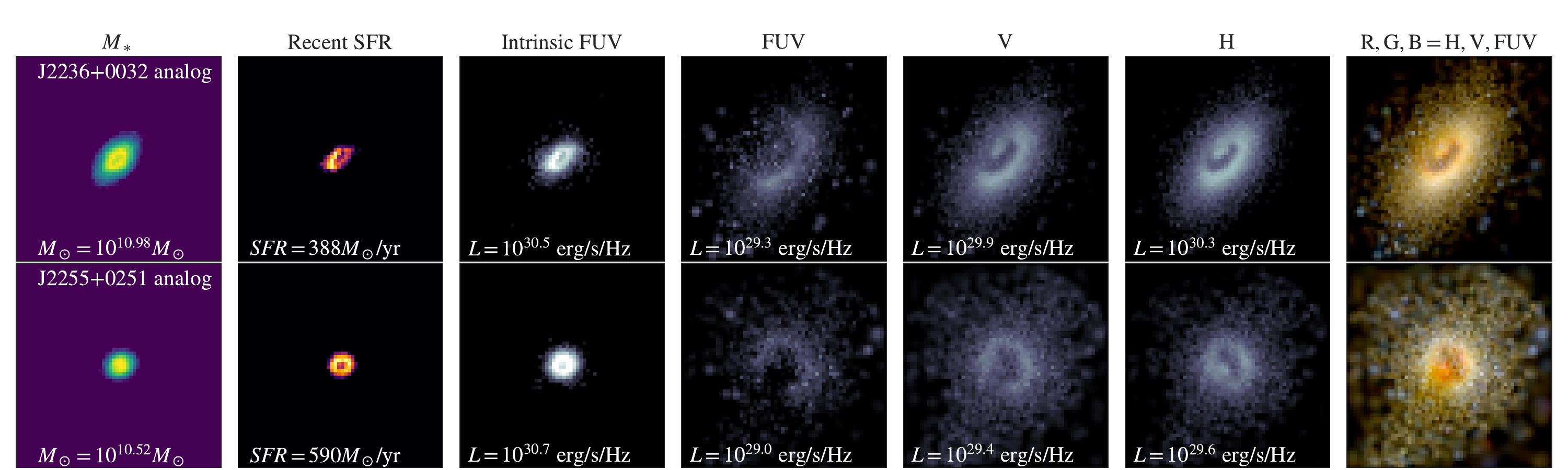}
    \caption{Rest frame images of BlueTide analogs. From left to right in each panel, we show the following i) stellar mass distribution in each pixel, ii) star formation rate of the z = 6.5 galaxy, iii) intrinsic far UV image without dust, iv-vi) far UV (FUV) images, visible (V), and near infrared (H) images with dust, and v) the stacked FUV, V, and H images.     \label{fig:restframe_bluetides}}
\end{figure*}

With the intrinsic properties simulated in BlueTides, we also create rest frame images of galaxies. These provide a glimpse into the evolution of detected hosts of the \citet{10.48550/arxiv.2211.14329} and other SHELLQs quasars that is unachievable with observations. We use \texttt{SynthObs} to generate rest frame images of the BlueTides analogs. These images do not have the observational effects of the other mock images, such as noise and PSF effects, and have a much higher resolution. We show the stellar mass distribution and recent star formation in Figure \ref{fig:restframe_bluetides}. We also show images in idealized top hat filters in the far UV, visible (V), and infrared (H) as well as the stacked RBG version. These images provide a higher resolution snapshot of the star particles in the BlueTides analog than the blurred redshifted images shown in Figures \ref{fig:mock_1} through \ref{fig:bright_analog}. The galaxies are notably compact in the left three panels in stellar mass, SFR, and intrinsic FUV. The intrinsic FUV panel traces the star formation and stellar mass. This is in line with results from \citet{marshall_2022_dust} which predicted that galaxies hosting quasars are more compact intrinsically relative to the entire BlueTides sample. When we include dust attenuation in the four right panels of Figure \ref{fig:restframe_bluetides}, we see a flatter emission profile as predicted in \citet{marshall_2022_dust}. This results in the appearance of an extended galaxy. The ring around the center of these galaxies shows a region where the dust is particularly strong. 

\subsection{Point source removal}
After the creation of mock images described in section \ref{subsec:mock_images_methods}, we compare the mock imaged BlueTides galaxies before and after the point source removal process to further increase the confidence that the \citet{10.48550/arxiv.2211.14329} hosts appear to be plausible detections. To do this, we perform quasar subtraction using the software \texttt{psfmc} on our model images, using a mock star image as our input PSF model. For our mock star image, we use the same model PSF from WebbPSF used to create our quasar image but with a different realization of the background and shot noise (see \citet{10.1093/mnras/stab1763}). This is an idealized case, where the PSF shape is known perfectly up to noise variation. While the most significant challenge in high-z quasar host detection is often the PSF mismatch error, our approach shows a more ideal case, which is useful for understanding the optimal outcome from these subtractions. Even with a perfect PSF shape, we will inevitably end up with some PSF subtraction error, which we begin to understand qualitatively here, and our aim is to see how this might appear in the \citet{10.48550/arxiv.2211.14329} host images.

We recover the properties of the host galaxy and quasar from the imaged galaxies using the software \texttt{psfmc} to do point source extraction. \texttt{psfmc} was written specifically for quasar subtraction unlike other algorithms that perform similar 2D surface brightness fitting such as \texttt{galfit} in \citet{2002AJ....124..266P}. \texttt{psfmc} uses the Affine Invariant Markov chain Monte Carlo (MCMC) ensemble sampler implemented in \texttt{emcee} \citep{Foreman_Mackey_2013}. The posterior sampled includes priors specified by the user and the following Gaussian joint likelihood for all pixels
\small
\begin{equation}
P(\rm image(pixel) | \theta) = \frac{1}{\sqrt{2 \pi \sigma(\rm pixel)^2}} \rm exp \left(\frac{(\rm image(pixel) - \rm I_{CM}(\rm pixel))^2}{2 \sigma(\rm pixel)^2}\right),
\end{equation}
\normalsize
where $\sigma(\rm pixel)$ includes the variance at that pixel in both the PSF and raw image and $\rm I_{CM}(\rm pixel)$ is the intensity of the convolved model at that particular pixel given some set of model parameters $\theta$. For each run of \texttt{psfmc}, we pass in a PSF generated by an imaged point source with a total flux of $6 \times 10^{5}~\rm nJy$ and convolved with the observed filter's mock PSF from WebbPSF using \texttt{SynthObs}.

The output of \texttt{psfmc} contains the posteriors of all parameters and the final galaxy image with the quasar subtracted. The intermediate images derived are the raw model and the convolved model. The raw model contains the highest probability galaxy and quasar surface brightness distribution. We convolve the raw model with the PSF to form the convolved model which is subtracted from the input observed image to determine the residuals. The third column of the panels labeled "model" in figures \ref{fig:mock_1} through \ref{fig:mock_4} show the convolved model.

We employ the sampling techniques from \texttt{psfmc} in the same filters as those observed in JWST. We model each quasar and its host as a point source and Sérsic profile defined galaxy system. We specify uniform priors for their positions, the quasar magnitude, half light semi-minor and major axis (the latter is taken as the half light radius), magnitude, and proper angle. We set the Sérsic index to 1 which is commonly done (i.e., in \citealt{10.48550/arxiv.2211.14329} and \citealt{yue2023eiger}) and has been shown to be consistent with simulated high-z galaxies \citep{article_sersic}. We sample the following free parameters with uniform priors: $m_{\rm filter, quasar}$, $m_{\rm filter, S\acute{e}rsic}$, $r_{\rm 1/2}$, $r_{\rm 1/2, b}$, and Sérsic angle. The uniform prior ranges are as follows for all Sérsic profile fitting and point source removal in this work:
\begin{enumerate}
    \item $m_{\rm filter, quasar}$ - [21, 25]
    \item $m_{\rm filter, S\acute{e}rsic}$ - [22, 26]
    \item $r_{\rm 1/2}$ [pixels] - [1, 16]
    \item $r_{\rm 1/2, b}$ [pixels] - [1, 11]
    \item Sérsic index - fixed at 1
    \item Sérsic angle [deg] - [0, 180]
\end{enumerate}
We initiate 64 walkers and discard the initial 200-400 samples to allow proper convergence of our sampler. The values of our results are drawn from the last 1k sample ensuring accurate parameter error estimation. The mock images are shown and compared qualitatively in the subsequent sections \ref{subsec:mock_images_actual} and \ref{subsec:mock_measured_props}.

\subsection{Mock Images of BlueTides Analogs}
\label{subsec:mock_images_actual}
% We simulate the JWST observation pipeline by imaging BlueTides galaxies with added quasars of the same magnitude to those observed in \citet{10.48550/arxiv.2211.14329}. This provides us with another quantitative avenue to compare the detected quasars to simulations. The BlueTides galaxy host for each quasar is a galaxy chosen such that it hosts a quasar with similar properties. This described in depth in section \ref{subsec:jwst_analog_selection}. 
We also generate realistic JWST NIRCam mock observations for each analog. This serves as a useful comparison as we are able to compare the exact BlueTides galaxy with and without the source extracted. This is evidently impossible with actual observations but allows us to test how accurately the point source subtraction reveals the underlying stellar light. 

\begin{figure*}
    \begin{subfigure}[t]{\textwidth}
    \includegraphics[width=\linewidth]{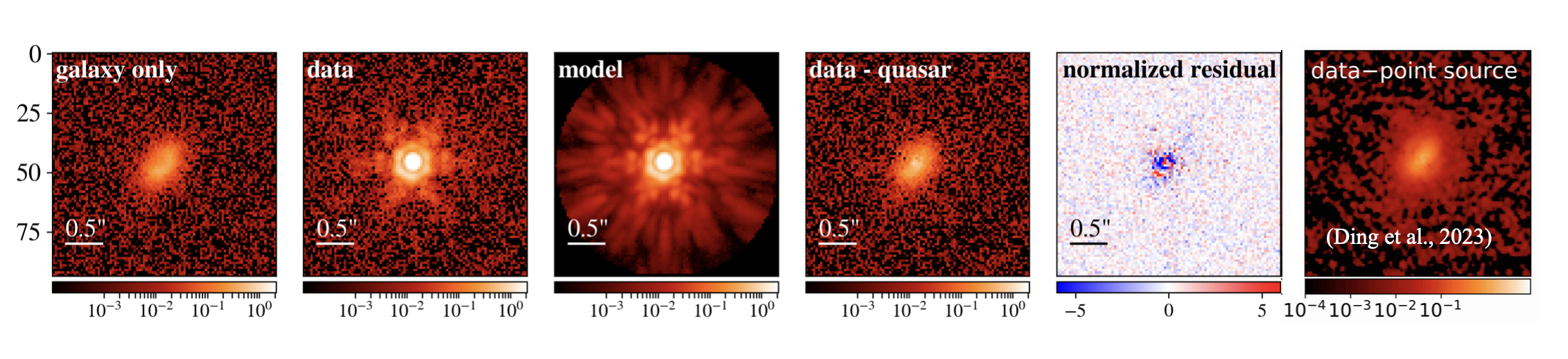}
    \caption{The BlueTides analog to J2236+0032 imaged in the F356W filter.}
    \label{fig:mock_1}
    \end{subfigure}
    \begin{subfigure}[t]{\textwidth}
    \includegraphics[width=\linewidth]{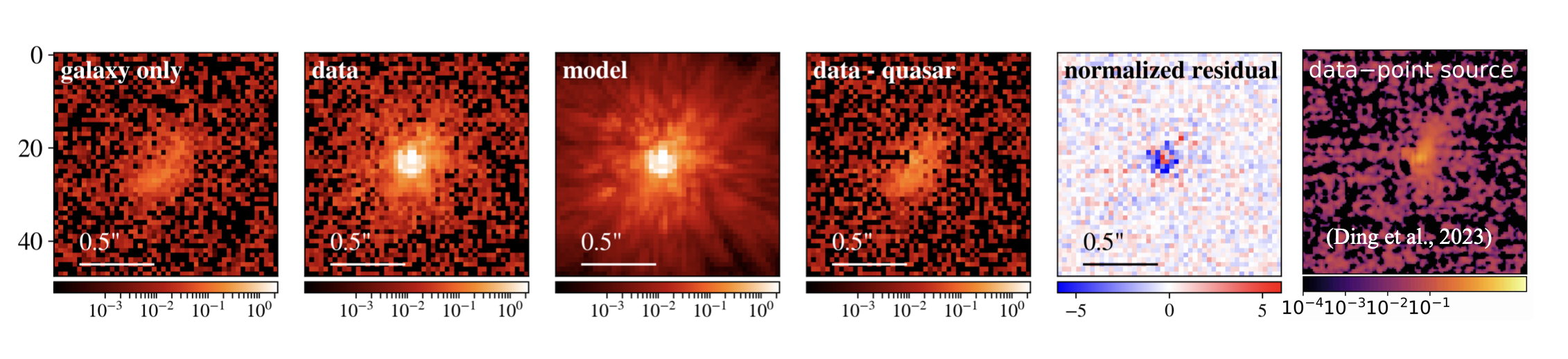}
    \caption{The BlueTides analog to J2236+0032 imaged in the F150W filter.}
    \label{fig:mock_2}
    \end{subfigure}
    \begin{subfigure}[t]{\textwidth}
    \includegraphics[width=\linewidth]{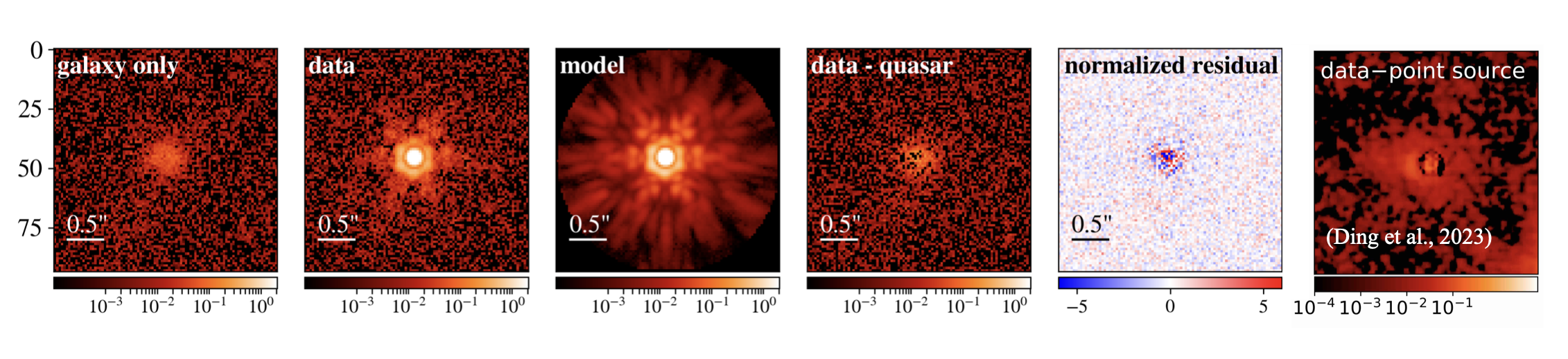}
    \caption{The BlueTides analog to J2255+0251 imaged in the F356W filter.}
    \label{fig:mock_3}
    \end{subfigure}
    \begin{subfigure}[t]{\textwidth}
    \includegraphics[width=\linewidth]{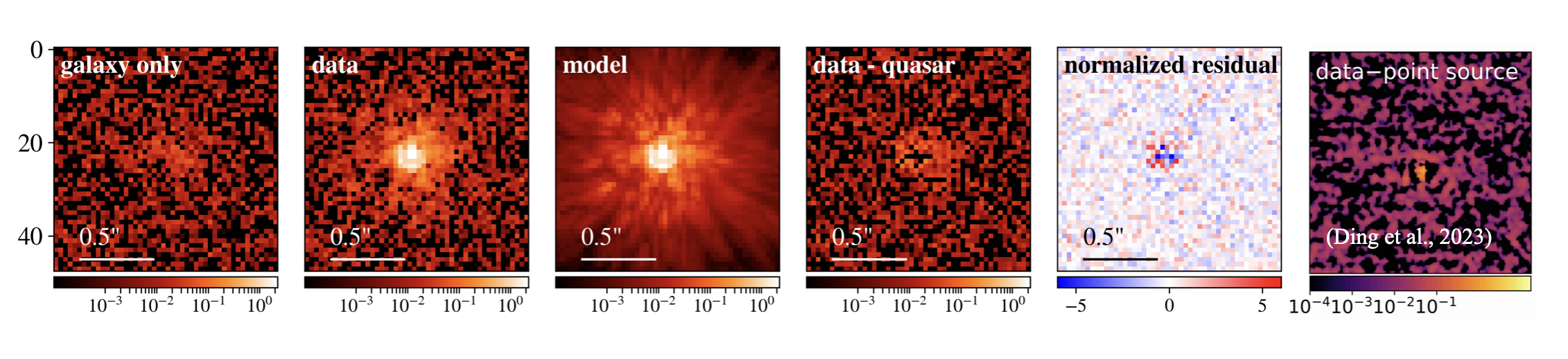}
    \caption{The BlueTides analog to J2255+0251 imaged in the F150W filter.}
    \label{fig:mock_4}
    \end{subfigure}
    \caption{Left to right of each panel: The host galaxy without the quasar imaged (very faint in F150W), the mock imaged quasar, the PSF convolved model which includes both the point source and galaxy, the data with the quasar extracted, the residual (data - model) normalized by the standard deviation of the residuals, and a copy of a portion of Figure 2 from \citet{10.48550/arxiv.2211.14329} showing the corresponding detected quasar host. Color bar units are MJy/str. Each image is a cut out from the full image and is equivalent in size to the rightmost plot in each panel with a size of 1.5" and 3" for F150W and F356W, respectively. The smaller size of F150W in \citet{10.48550/arxiv.2211.14329} is due to supersampling in both filters such that the pixel size in each filter varies by a factor of two (0.0315 arcsec/pixel for F356W and 0.0153 arcsec/pixel for F150W). Our pixels size is 0.031 arcsec/pixel in both filters. The corresponding pixel numbers between our images and the \citet{10.48550/arxiv.2211.14329} are not equivalent.  \label{fig:all_mock}}
\end{figure*}

\begin{figure*}
    \begin{subfigure}[t]{\textwidth}
    \includegraphics[width=\linewidth]{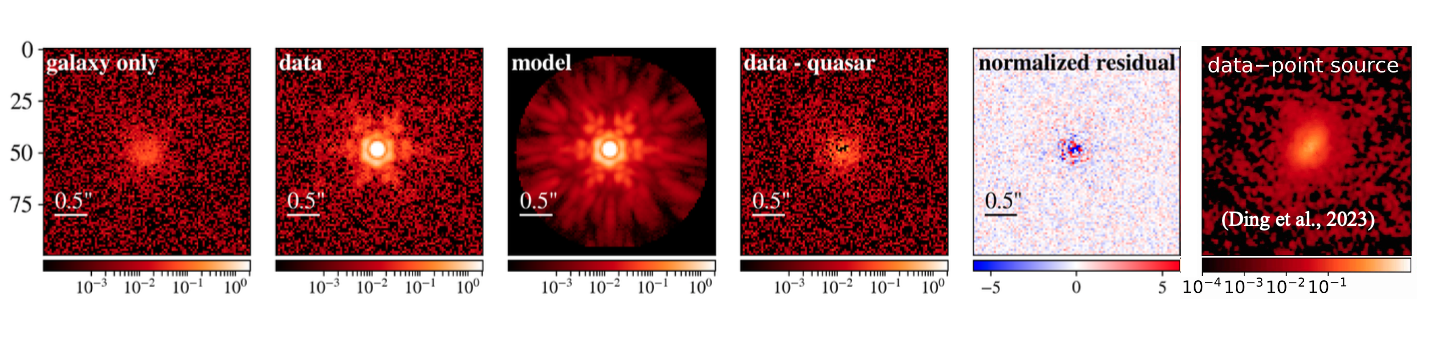}
    \caption{.}
    \label{fig:dim_analog}
    \end{subfigure}
    \begin{subfigure}[t]{\textwidth}
    \includegraphics[width=\linewidth]{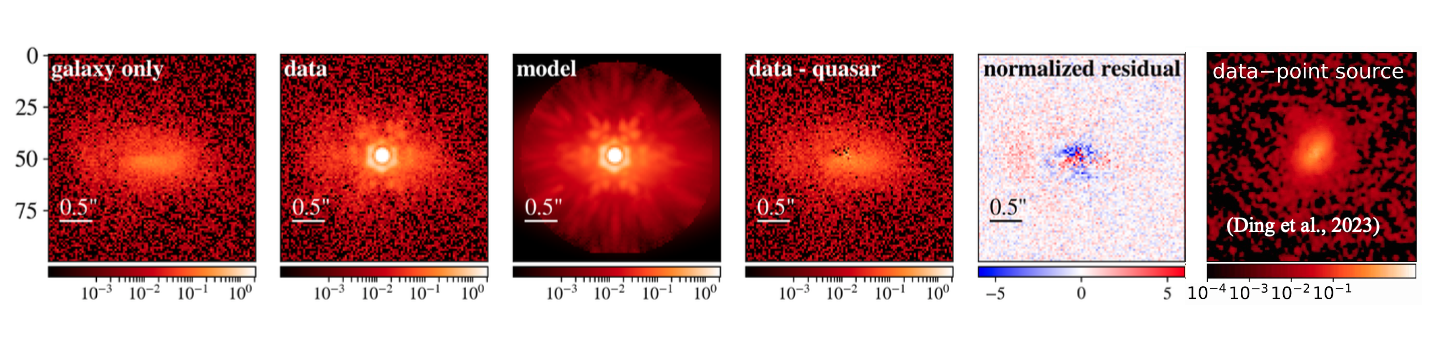}
    \caption{}
    \label{fig:bright_analog}
    \end{subfigure}
    \caption{Same panels as Figure \ref{fig:all_mock}. Top: The closest analog to J2236+0032 with a moderately successful point source subtraction in the F356W filter. This analog is within 20$\%$ of the bolometric luminosity of J2236+0032 rather than within 50$\%$ like the analog presented in this work. Bottom: Brightest galaxy among the first 100 BlueTides galaxies with the most highly accreting black holes. All color bar units are MJy/str. Note that each mock image is 3" in width and roughly the same size as the \citet{10.48550/arxiv.2211.14329} .}
\end{figure*}

In Figures \ref{fig:mock_1} through \ref{fig:bright_analog}, we show six panels of our mock observed images, depicting I) the BlueTides galaxy without a quasar imaged, II) the same BlueTides galaxy with a quasar imaged, III) the convolved model from \texttt{psfmc} described in section \ref{subsec:mock_images_methods}, IV) the host galaxy image with the quasar subtracted via \texttt{psfmc}, V) a normalized residual between the data and point source subtracted image from \texttt{psfmc} (i.e., (data - model) / $\sigma_{\rm data-model}$), and VI) the detected host JWST image from \citet{10.48550/arxiv.2211.14329}. With \texttt{psfmc}, we are successful in extracting the 2D galactic brightness distribution and quasar parameters for all of our imaged BlueTides galaxies. This is promising as the analogs are particularly faint in the F150W filter. The actual host (galaxy only image) versus quasar subtracted (data - quasar image) all recover the underlying host with moderate visible error. In Figures \ref{fig:mock_2} and \ref{fig:mock_3}, the center of the host is missing flux. Each quasar subtracted host also features subtle to significant morphological changes from the actual host image. Through the normalized residuals in the fourth panel, we can quantitatively compare the data + quasar versus galaxy only images which ideally would be the same. There is a slight underextraction and overextraction around the core area containing the point source. This is an artifact of the point source removal and should be considered with point source subtraction on real images. 

In Figures \ref{fig:mock_1} through \ref{fig:mock_3}, the J2236+0032 and J2255+0251 host analogs are similar to the detected hosts in \citet{10.48550/arxiv.2211.14329} in all filters. In Figure \ref{fig:mock_4}, the J2255+0251 host analog detection in F150W is shown alongside the non-detection in \citet{10.48550/arxiv.2211.14329}. Notably, this is also the dimmest host analog detection in our sample.

In Figure \ref{fig:dim_analog}, we show the mock images from a \textit{dim analog} matching properties of J2236+0032. This analog is only slightly different in magnitude and black hole bolometric luminosity compared to the analog chosen in Figures \ref{fig:mock_1} through Figure \ref{fig:mock_2}. The dim analog hosts a black hole within 20$\%$ versus 50$\%$ of J2236+0032's bolometric luminosity and is approximately 1 mag fainter than the standard analog we chose. This analog is detected but with a less visibly accurate galaxy reconstruction compared to the original analog shown in Figure \ref{fig:mock_1}. This is seen qualitatively by comparing the galaxy and data - quasar panels where the dimmer galaxy has a larger corrupted quasar core.

In Figure \ref{fig:bright_analog}, we show the brightest galaxy in our BlueTides sample that hosts one of the 100 most luminous black holes to show the point source removal on a different galaxy morphology. This galaxy looks more elongated than any of the other analogs and may be undergoing a merger. The elongated morphology of this galaxy hosting one of the most luminous black holes in BlueTides supports further host galaxy comparison to those detected in \citet{10.48550/arxiv.2211.14329} and \citet{yue2023eiger}. The size and luminosity of the galaxy seems to play a significant role in point source subtraction and will be explored in a future work.

\begin{figure*}
    \includegraphics[width=0.6\linewidth]{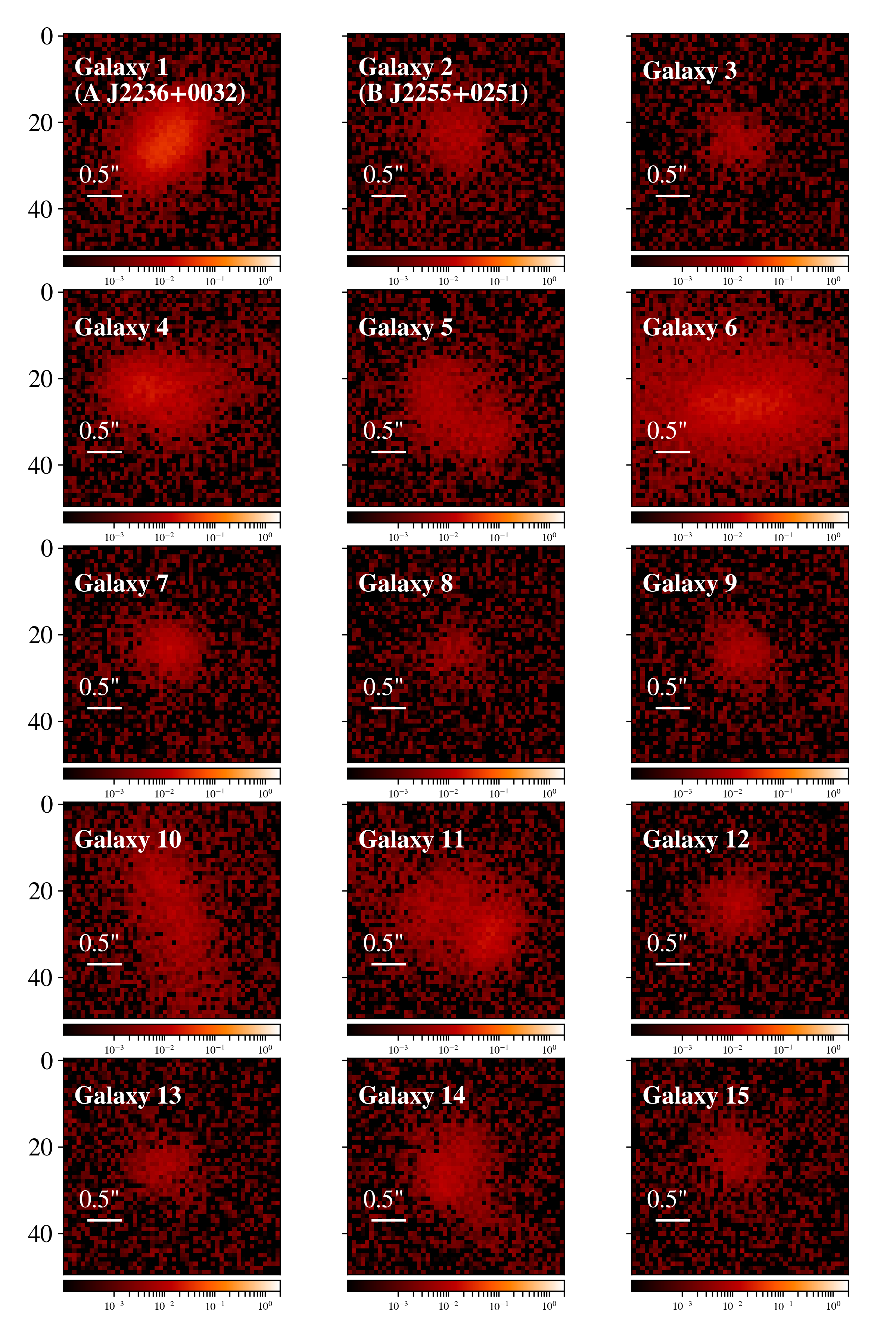}
    \caption{Mock images in the F356W filter of 15 galaxies in BlueTides with similar bolometric luminosities to J2236+0032 or J2255+0251. Half of the images show compact morphology similar to the observed high-z quasar hosts detected in \citet{10.48550/arxiv.2211.14329}. These images are smoothed prior to imaging. Color bar units are MJy/str.}
    \label{fig:just_galaxies}
\end{figure*}

In Figure \ref{fig:just_galaxies}, we show a wide array of \textit{mock images} of BlueTides galaxies with similar bolometric luminosities to J2236+0032 and J2255+0251 in the F356W filter for comparison. At z = 6.5, these galaxies look substantially more diffuse than those detected in \citet{10.48550/arxiv.2211.14329} with some showing evidence of mergers. The EIGER galaxy hosts which contain quasars with higher luminosity resemble these larger black hole hosts. The J0148+0600 and J159-02 detected hosts show evidence of companion galaxies, and J1120+0641 is notably more diffuse. This supports further statistical studies on physical properties of host galaxies of high-z quasars to constrain evolutionary models.

\subsection{Measured Properties in Mock Images}
\label{subsec:mock_measured_props}
Qualitatively, the imaged galaxies look similar to the detected galaxies in \citet{10.48550/arxiv.2211.14329}, with similar shapes and compactness. However for most of these mock images, our point source subtraction algorithm described in section \ref{subsec:mock_images_methods} creates moderate to large errors in predicted radii and magnitude when extracting a point source. We compare the Sérsic radii and magnitudes quantitatively for the imaged galaxy with and without the point source. We extract both Sérsic radii and magnitudes as defined in equation \ref{eq:sersic} using the MCMC framework in \texttt{psfmc} to maintain consistency. We tabulate the mean and one sigma value from our sampled posteriors for each Sérsic radius in Table \ref{tab:peaks} for both the original galaxy before adding the point source and the radius estimated after point source subtraction. We find that the results do not always agree to one sigma for both the original and point source subtracted galaxies indicating that point source subtraction is introducing bias. 

The radii have a large discrepancy between simulated and observed values. We note that the largest bias in radius is found for the J2236+0032 analog in F150W shown in Figure \ref{fig:mock_2} with the radii differing by about 0.4 pkpc and a percent difference of $\sim 30\%$. The radii are always underestimated from the actual value in our analog sample shown in Table \ref{tab:peaks}. However, the radius of the \textit{brightest galaxy} is recovered relatively accurately. This indicates that there is a larger error for smaller hosts that are more hidden behind the corrupted quasar core. This exemplifies the caution needed when using measured radii values from high-z hosts. We also compare these radii to the dust attenuated half light radii from \citet{marshall_2022_dust} as described in section \ref{sec:radii}. These radii are on the same order as the radii found through the other methods but do not match even those radii measured without an added quasar. However, this is expected as the radii from \citet{marshall_2022_dust} are not from a Sérsic profile fit.

We also measure the galaxy magnitude for the BlueTides analog with and without the point source removal process. We find that the magnitudes are predicted more accurately with each measured magnitude falling within approximately one sigma of the pre-point source removal value. Similarly to the radii measurements, we find that the post point source removal magnitude measurement is underestimated approximately $50\%$ of the time compared to the actual Sérsic magnitude of the host.
% \red{The largest bias in magnitude is also found in the J2236+0032 analog in F150W with a percent difference of $5\%$ and magnitude difference of 0.137 mag. When we compare the magnitudes tabulated in Table \ref{tab:quasar_mags} found through mock observations in \texttt{SynthObs} and those found with \texttt{psfmc} in Table \ref{tab:peaks}, the J2236+0032 magnitudes are within 0.5 mag of each other, but the J2255+0251 magnitudes differ by approximately 1 mag. The magnitudes found with \texttt{SynthObs} are always lower than those found with \texttt{psfmc}. This is due to the flux from every pixel in the \texttt{SynthObs} mock image being included in the total flux. \texttt{psfmc} confines the magnitude to the galaxy radius when fitting the Sérsic profile which results in a fainter magnitude overall.} 

Lastly, we compare the magnitude and radius pre- and post-point source removal for the brightest BlueTides galaxy within the first 100 most massive black holes and imaged in Figure \ref{fig:bright_analog}. As seen in Table \ref{tab:peaks}, the radius is predicted the most accurately for this galaxy. Similar to the other quasar analogs, the observed magnitude shows a slight underestimation post point source removal. 

We note that the errors we find for our analogs are smaller than those seen for the observed quasars and their hosts in \citet{10.48550/arxiv.2211.14329}. Our errors include errors from our more idealized point source removal, as well as the noise in the image. In the observations in \citet{10.48550/arxiv.2211.14329}, the uncertainty also includes larger systematic errors due to PSF mismatching and other effects caused by the real telescope imaging process.
% \red{It should be clarified that the reason for the authors' error being smaller than in Ding et al. (2023) is not solely due to using galight errors. The errors of these analogs stem from random error resulting from one image simulation seed. In Ding's work, systematic errors are estimated by considering factors such as PSF mismatch and different modeling technique strategies, (such as pixel supersampling factors in the models).}

Although we have not conducted a robust statistical study in this work, we see a trend of more accurately predicted galactic properties for brighter galaxies. With our mock imaging pipeline and point source removal algorithm, we can use BlueTides to quantify bias and make estimates for the over or underestimation of quasar host properties more generally. This will be the subject of a future paper using the large sample of BlueTides galaxies.

\begin{table*}
\centering
\caption{Mean and one sigma error of sampled values for each Sérsic radius and magnitude for BlueTides galaxies. Values are shown with and without the added quasars for each mock imaged analog. We show both quasar analogs, the dimmer J2236 analog, and the brightest galaxy hosting one of the first 100 most luminous black holes. Each Sérsic radius and magnitude was generated using uniform priors and \texttt{psfmc}. All errors shown on the properties are one sigma. The BlueTides galaxy indicates which of the two galaxies was used in each image. The right column shows the dust half light radius described in section \ref{sec:radii} from \citet{marshall_2022_dust}. \label{tab:peaks}}
\resizebox{\textwidth}{!}{
\begin{tabular}{lllllllr}
\hline
Quasar Analog              & Filter               & BlueTides Galaxy     & $m_{\rm S\acute{e}rsic}$ without quasar [mag] & $m_{\rm S\acute{e}rsic}$ [mag] & $R_{\rm 1/2}$ without quasar {[}kpc{]} & $R_{\rm 1/2}$ [kpc]  & $R_{\rm 1/2}$ [kpc] \citep{marshall_2022_dust} \\ \hline
 J2255 &  F356W &               A &  25.009 $\pm$ 0.063 &  25.047 $\pm$ 0.073 &  1.318 $\pm$ 0.093 &  1.155 $\pm$ 0.171 & 1.4 \\ \hline
 J2255 &  F150W &               A &  25.879 $\pm$ 0.079 &  25.892 $\pm$ 0.079 &  1.673 $\pm$ 0.187 &   1.326 $\pm$ 0.25 & 1.4 \\ \hline
 J2236 &  F356W &               B &  23.718 $\pm$ 0.041 &  23.579 $\pm$ 0.022 &  1.209 $\pm$ 0.043 &  0.898 $\pm$ 0.055 & 1.7 \\ \hline
 J2236 &  F150W &               B &  25.034 $\pm$ 0.051 &   24.902 $\pm$ 0.05 &    1.324 $\pm$ 0.1 &   0.96 $\pm$ 0.074 & 1.7 \\ \hline
(Not an analog) &  F356W &  Brightest galaxy (within 100 brightest BHs) & 23.134 $\pm$ 0.014 &  23.062 $\pm$ 0.019 &  2.812 $\pm$ 0.004 &  2.803 $\pm$ 0.038 &           1.3 \\ \hline
 J2236 &  F356W &      Dim analog &  24.737 $\pm$ 0.059 &  24.727 $\pm$ 0.056 &  1.522 $\pm$ 0.093 &  1.223 $\pm$ 0.114 & 1.7 \\ \hline

\end{tabular}}
\end{table*}

\section{Detection predictions for J2255+0251 and J2236+0032}
\label{sec:predict}
\begin{figure}
\centering
    \includegraphics[width=0.8\columnwidth]{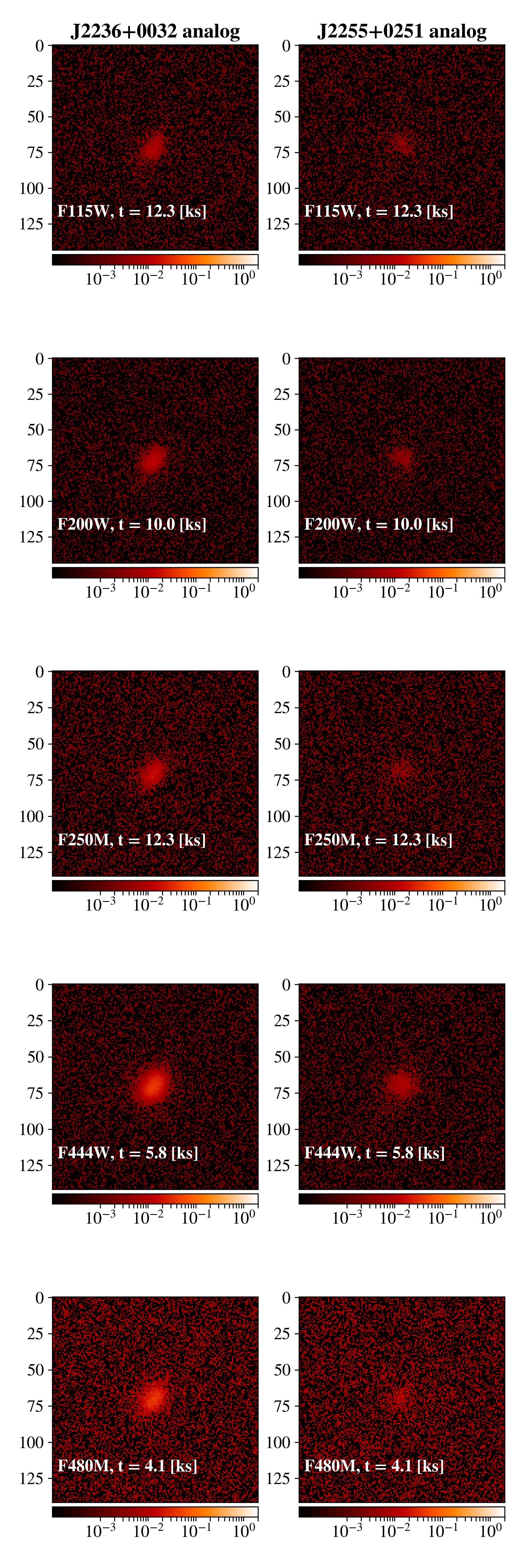}
    \caption{Mock imaged BlueTides analogs in the filters and exposure times described in Table \ref{tab:jwst_proposal_2}. These images allow us to predict positive detection of the host in all filters and exposure times but are notably fainter for J2255+0251. These images are smoothed prior to imaging.}
    \label{fig:proposal_filters}
\end{figure}

\begin{figure}
\centering
    \includegraphics[width=0.9\columnwidth]{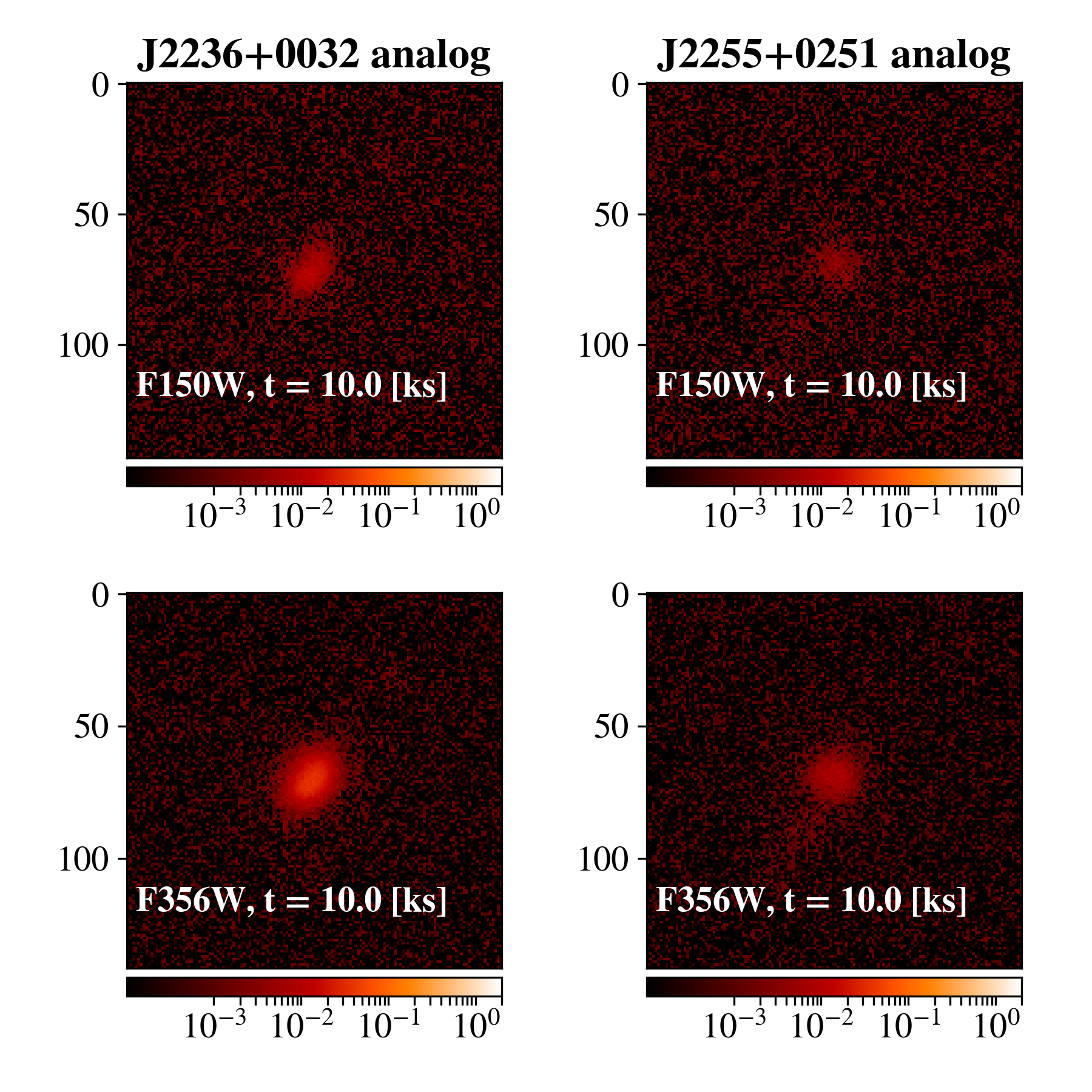}
    \caption{Same as Figure \ref{fig:proposal_filters} but for a mock 10ks exposure of the F356W and F150W filters which are imaged with 3.1ks exposure times in \citet{10.48550/arxiv.2211.14329}. This filter and exposure time is not included in followup proposals but shows a prediction for a longer exposure time with Figure \ref{fig:proposal_filters}. Color bar units are MJy/str. These images are smoothed prior to imaging.}
    \label{fig:longer}
\end{figure}

Follow up observations with JWST of J2255+0251 and J2236+0032 (detected in both F150W and F356W by JWST) will allow further constraints on the properties of their hosts. In JWST Cycle 2 GO $\#$ 3859 (PI: Onoue), these quasars will be followed up in five NIRCam filters. In Table \ref{tab:jwst_proposal_2}, the filters and exposure times for this proposal are shown. In Figure \ref{fig:proposal_filters}, we show the mock images of galaxies without an added quasar for these filters and exposure times. We also extend the exposure time of the original filters (F356W and F150W) to 10ks in Figure \ref{fig:longer} for comparison to the 3.1ks exposure times shown in figures \ref{fig:mock_1} through \ref{fig:mock_4}. We supersample in all filters with wavelengths larger than $2 \mu \rm m$ (filters above F200W) to maintain consistency with our mock imaging in section \ref{sec:mock_images}. 

The analogs are visible in most filters but are notably fainter in the medium band filters (F250M and F480M) and at filters including wavelengths below 2 microns. We expect that a host detection for the more compact and faint host of J2255+0251 in GO $\#$3859 will be very difficult in F115W, F200W, F250M, and F480M. The brighter and slightly elongated host J2236+0032 may be relatively easier to detect in all filters. In Figure \ref{fig:longer} for F150W, the analog of J2255+0251 is still extremely faint which may indicate host detection in this filter is impossible.

Although a simulated version of the true hosts of J2255+0251 and J2236+0032, our detection predictions will be useful in future JWST observations in filter and exposure time selection of high-z quasars and as a test of the predictive power of BlueTides.

\begin{table}
\centering
\caption{Overview of the followup filters and exposure times for J2236+0032 in \citet{10.48550/arxiv.2211.14329}.}
\begin{tabular}{cc}
\hline
Filter  & Exposure Time [s] \\ \hline
F115W                 & 12369 \\ \hline
F200W                 & 9964\\ \hline
F250M                 & 12369 \\ \hline 
F444W                 & 5841 \\ \hline
F480M &  4123\\ \hline
\label{tab:jwst_proposal_2}
\end{tabular}
\end{table}

% made in  paper_1_bluetides/helper scripts/panel_8_proposal

\section{Conclusions}
\label{sec:conclusions}
The launch of JWST has opened a new era in high-z quasar observation with J2255+0251 and J2236+0032 recently becoming the first high-z quasars with host galaxy detections \citep{10.48550/arxiv.2211.14329}. Soon after, three more quasar hosts were detected \citep{yue2023eiger}. To understand the properties of high-z quasars, we compare the recently detected quasar hosts to systems hosting black holes with similar bolometric luminosities in the BlueTides simulation. To summarize, we find the following results:
\begin{enumerate}
    \item We find the measured host stellar mass, radii, and black hole mass of J2255+0251 and J2236+0032 are consistent with similar BlueTides quasars. The expected star formation rates of J2255+0251 and J2236+0032 from comparisons with quasars of similar luminosities is between $\log_{10} \rm SFR \sim 2-3 M_{\odot}/\rm yr$. We also set a lower limit of $10^{2.75}$ $M_{\odot}/ \rm yr$ for the EIGER quasars.
    \item We find that the newly detected EIGER quasar hosts \citet{yue2023eiger} do not match the BlueTides simulation host predicted properties and thus do not align with the low-z Magorrian relation. However, the bolometric luminosities of these quasars are at the bright end of the BlueTides sample (See the quasar luminosity functions in Figure \ref{fig:quasar_pdf}). It is not yet clear whether the disagreement is due to an observational bias or a bias owing to the finite simulation volume where we do not have a statistical comparative sample.
    \item With the criteria described in section \ref{subsec:jwst_analog_selection}, we select two analogs similar to J2255+0251 or J2236+0032 in stellar mass, black hole mass, and galactic radius. We compare the BlueTides analogs to the actual JWST detected quasar hosts and find the photometric galaxy magnitudes match closely between simulation and observation. We use mock imaging and a point source subtraction algorithm to make qualitative comparisons and begin to understand the effects of observational biases with these two analogs. We find that the mock images of BlueTides analogs are qualitatively similar to the detected hosts in \citet{10.48550/arxiv.2211.14329}. However, the host detection remains difficult. This supports the lack of quasar host detections in \citet{stone2023undermassive}. We do detect all hosts, but note that the success is highly dependent on the luminosity of the host galaxy. When we determine the half light radius and magnitudes for mock images of galaxies, we find that the recovered magnitudes agree mostly to within one sigma. The radii have a larger discrepancy between actual and post point source removal values with a maximum percent difference of $\sim 30\%$ or about 0.4 pkpc. The point source removal process consistently underestimates the radii in the galaxies tested in Table \ref{tab:peaks}. This motivates future work to quantify these biases. 
    \item We make predictions for the detections of J2236+0032 and J2255+0251 in upcoming JWST observations. We predict that longer exposure times and longer wavelength filters will provide the highest probability for successful host detections but the J2255+0251 host may remain elusive even with a longer exposure time in the F150W filter.
\end{enumerate}

Our matched comparisons to BlueTides show the utility of high-z simulations to study high-z quasar hosts. Despite being tuned to low-z observations, BlueTides is still able to accurately predict high-z galaxy and black hole properties. In addition to confirming agreement of observed and simulated properties, BlueTides will serve as a tool to infer the properties of high-z quasar hosts from observations. 

\section*{Data Availability}
Data from the BlueTides simulation is available at \url{http://bluetides.psc.edu}. Other data generated in this work will be shared on reasonable request to the corresponding author. The reproduced \citet{10.48550/arxiv.2211.14329} images in this article were provided by Springer Nature with permission.

\section*{Acknowledgements}
We thank the anonymous referee for their comments which greatly helped to improve this work.

We are grateful to Xuheng Ding and the rest of the SHELLQs team for helpful email communication and allowing the use of their JWST images in this paper. We also thank Mira Mechtley who wrote the point source removal software, \texttt{psfmc}. S.B. thanks Chris Lovell, Yuxiang Qin and Xuejian Shen for helpful comments or code that contributed to this paper. 

Parts of this research were supported by the Australian Research Council Centre of Excellence for All Sky Astrophysics in 3 Dimensions (ASTRO 3D), through project number CE170100013. S.B. acknowledges the support of an Australian Government Research Training Program (RTP) Scholarship. MAM acknowledges the support of a National Research Council of Canada Plaskett Fellowship. This work was partially performed on the OzSTAR national facility at Swinburne University of Technology. OzSTAR is funded by Swinburne University of Technology and the National Collaborative Research Infrastructure Strategy (NCRIS). This research made use of Python packages NumPy \citep{harris2020array}, Matplotlib \citep{Hunter:2007}, AstroPy \citep{astropy:2013, astropy:2018, astropy:2022}, SciPy \citep{2020SciPy-NMeth}, emcee \citep{Foreman_Mackey_2013}, and corner \citep{corner}.


\bibliographystyle{mnras}
\bibliography{main} % if your bibtex file is called example.bib

% Don't change these lines
\bsp	% typesetting comment
\label{lastpage}
\end{document}